\documentclass[10pt,conference]{IEEEtran}
\usepackage{cite}
\usepackage{amsmath,amssymb,amsfonts}
\usepackage{algorithmic}
\usepackage{graphicx}
\usepackage{textcomp}
\usepackage{xcolor}
\usepackage[hyphens]{url}
\usepackage{fancyhdr}
\usepackage{hyperref}
\usepackage{xspace}
\usepackage{tikz}
\newcommand*\circled[1]{\tikz[baseline=(char.base)]{
            \node[shape=circle,draw,inner sep=0.5pt] (char) {#1};}}

\newcommand{\hpcayear}{2026}

\newcommand{\sysname}{V-Rex\xspace}
\newcommand{\sysnames}{V-Rex's\xspace}
\newcommand{\algoname}{ReSV\xspace}
\newcommand{\hpcasubmissionnumber}{515}

\title{V-Rex: Real-Time Streaming Video LLM Acceleration via Dynamic KV Cache Retrieval}

\def\hpcacameraready{} 

\newcommand\hpcaauthors{Donghyuk Kim, Sejeong Yang, Wonjin Shin and Joo-Young Kim}
\newcommand\hpcaaffiliation{\textit{KAIST}\\ \textit{Daejeon, Republic of Korea}}
\newcommand\hpcaemail{\{kar02040, 02yangsj, 2swj1202, jooyoung1203\}@kaist.ac.kr}

\author{
  \ifdefined\hpcacameraready
    \IEEEauthorblockN{\hpcaauthors{}}
      \IEEEauthorblockA{
        \hpcaaffiliation{} \\
        \hpcaemail{}
      }
  \else
    \IEEEauthorblockN{\normalsize{HPCA \hpcayear{} Submission
      \textbf{\#\hpcasubmissionnumber{}}} \\
      \IEEEauthorblockA{
        Confidential Draft \\
        Do NOT Distribute!!
      }
    }
  \fi 
}

\fancypagestyle{camerareadyfirstpage}{%
  \fancyhead{}
  
  \fancyhead[C]{
    \ifdefined\aeopen
    \parbox[][12mm][t]{13.5cm}{\hpcayear{} IEEE International Symposium on High-Performance Computer Architecture (HPCA)}    
    \else
      \ifdefined\aereviewed
      \parbox[][12mm][t]{13.5cm}{\hpcayear{} IEEE International Symposium on High-Performance Computer Architecture (HPCA)}
      \else
      \ifdefined\aereproduced
      \parbox[][12mm][t]{13.5cm}{\hpcayear{} IEEE International Symposium on High-Performance Computer Architecture (HPCA)}
      \else
      \parbox[][0mm][t]{13.5cm}{\hpcayear{} IEEE International Symposium on High-Performance Computer Architecture (HPCA)}
    \fi 
    \fi 
    \fi 
    \ifdefined\aeopen 
      \includegraphics[width=12mm,height=12mm]{ae-badges/open-research-objects.pdf}
    \fi 
    \ifdefined\aereviewed
      \includegraphics[width=12mm,height=12mm]{ae-badges/research-objects-reviewed.pdf}
    \fi 
    \ifdefined\aereproduced
      \includegraphics[width=12mm,height=12mm]{ae-badges/results-reproduced.pdf}
    \fi
  }
  \fancyfoot[C]{}
}
\begin{document}
\maketitle

\ifdefined\hpcacameraready 
  \thispagestyle{camerareadyfirstpage}
  \pagestyle{empty}
\else
  \thispagestyle{plain}
  \pagestyle{plain}
\fi

\newcommand{\hpcaheight}{0mm}
\ifdefined\eaopen
\renewcommand{\hpcaheight}{12mm}
\fi


\begin{abstract}

Streaming video large language models (LLMs) are increasingly used for real-time multimodal tasks such as video captioning, question answering, conversational agents, and augmented reality. However, these models face fundamental memory and computational challenges because their key-value (KV) caches grow substantially with continuous streaming video input. This process requires an iterative prefill stage, which is a unique feature of streaming video LLMs. Prior works reduce excessive cache overhead by utilizing the KV cache retrieval algorithm, which offloads the full KV cache to CPU memory or storage, then selectively fetches the most relevant entries. Nevertheless, due to its iterative prefill stage, they suffer from significant limitations, including extensive computation, substantial data transfer, and degradation in accuracy. Crucially, this issue is exacerbated for edge deployment, which is the primary target for these models. The memory footprint exceeds the memory capacity within minutes of video streams, making low-latency, energy-efficient inference infeasible.

In this work, we propose \sysname, the first software-hardware co-designed accelerator that comprehensively addresses both algorithmic and hardware bottlenecks in streaming video LLM inference. At its core, \sysname introduces \algoname, a training-free dynamic KV cache retrieval algorithm. \algoname exploits temporal and spatial similarity-based token clustering to reduce excessive KV cache memory across video frames, and dynamically adjusts token selection per transformer layer and attention head to minimize the number of selected tokens. To fully realize these algorithmic benefits, \sysname offers a compact, low-latency hardware accelerator with a dynamic KV cache retrieval engine (DRE), featuring bit-level and early-exit based computing units, as well as hierarchical KV cache memory management. Evaluated on COIN benchmarks, \sysname achieves unprecedented real-time of 3.9-8.3 FPS and energy-efficient streaming video LLM inference on edge deployment with negligible accuracy loss. While DRE only accounts for 2.2\% power and 2.0\% area, the system delivers 1.9-19.7$\times$ speedup and 3.1-18.5$\times$ energy efficiency improvements over AGX Orin GPU. This work is the first to comprehensively tackle KV cache retrieval across algorithm and hardware, enabling real-time streaming video LLM inference on resource-constrained edge devices, with clear potential for scalable deployment in large-scale server environments.
\end{abstract}

\section{Introduction}
\label{sec_introduction}

Recently, the demand for artificial intelligence that can understand and interpret various modalities (i.e., text, image, video, and speech) and respond to inquiries has been a driving force in machine learning research. As a result, large multimodal models (LMM)~\cite{yin2024survey} have emerged as promising solutions in various AI industries. Notably, streaming video large language models (LLMs) have gained significant attention for their ability to jointly comprehend the video and text modalities in real-time. Streaming video LLMs demonstrate a wide range of tasks, including video captioning, question answering, conversational agents, and augmented reality.~\cite{chen2024videollm, xiong2025streaming, 10.1145/3706598.3714224}. Unlike offline video LLMs~\cite{tang2025video, shen2024longvu, zhang2023video}, it processes real-time video streams and responds to users’ inquiries, which primarily runs on edge devices. Due to continuous video input requiring an iterative process of video frames, computation and memory usage scale substantially. It causes the key-value (KV) caches to rapidly exceed the GPU memory capacity, and processing long video streams in real-time becomes impractical.

Existing KV cache optimizations are fundamentally mismatched for streaming and interactive workloads. Destructive methods, such as pruning~\cite{wu2024videollm}, compression~\cite{10.5555/3737916.3742359, kim2025infinipot, tu2024vl}, and quantization~\cite{tao2025plug, li2025commvq, liu2024kivi, kim2025oaken} risk permanently discarding information that, while irrelevant to the current query, may be essential for future ones, disrupting conversational continuity. A more promising approach, KV cache retrieval~\cite{lee2024infinigen, di2025streaming, ning2025livevlm}, avoids this issue by offloading the full cache to CPU memory or storage and fetching only relevant tokens, thereby reducing GPU memory usage while maintaining coherent responses for more extended input sequences. Although effective in reducing memory usage, they rely on bandwidth-limited links such as PCIe (4–32 GB/s), which are far slower than GPU memory bandwidth (1–2 TB/s). Thus, selective retrieval is necessary to avoid severe data transfer. 

However, current retrieval algorithms, designed for the text generation stage, perform poorly under the iterative prefill stage of streaming video. Moreover, their reliance on fixed top-k selection, which is a computationally regular and GPU-friendly primitive, introduces algorithmic inefficiencies. This static strategy ignores the highly variable importance of tokens across transformer layers and attention heads~\cite{tu2024vl, yang2024pyramidinfer, ge2023model}. Enforcing a fixed-k policy prioritizes hardware convenience over the algorithm's need, leading to systemic inefficiencies: either over-fetching redundant tokens, wasting PCIe bandwidth and energy, or under-fetching critical ones, degrading accuracy. Addressing this challenge requires more than an algorithmic tweak. It demands a new hardware optimization.

\begin{figure}[t]
\centering
\includegraphics[width=0.49\textwidth]{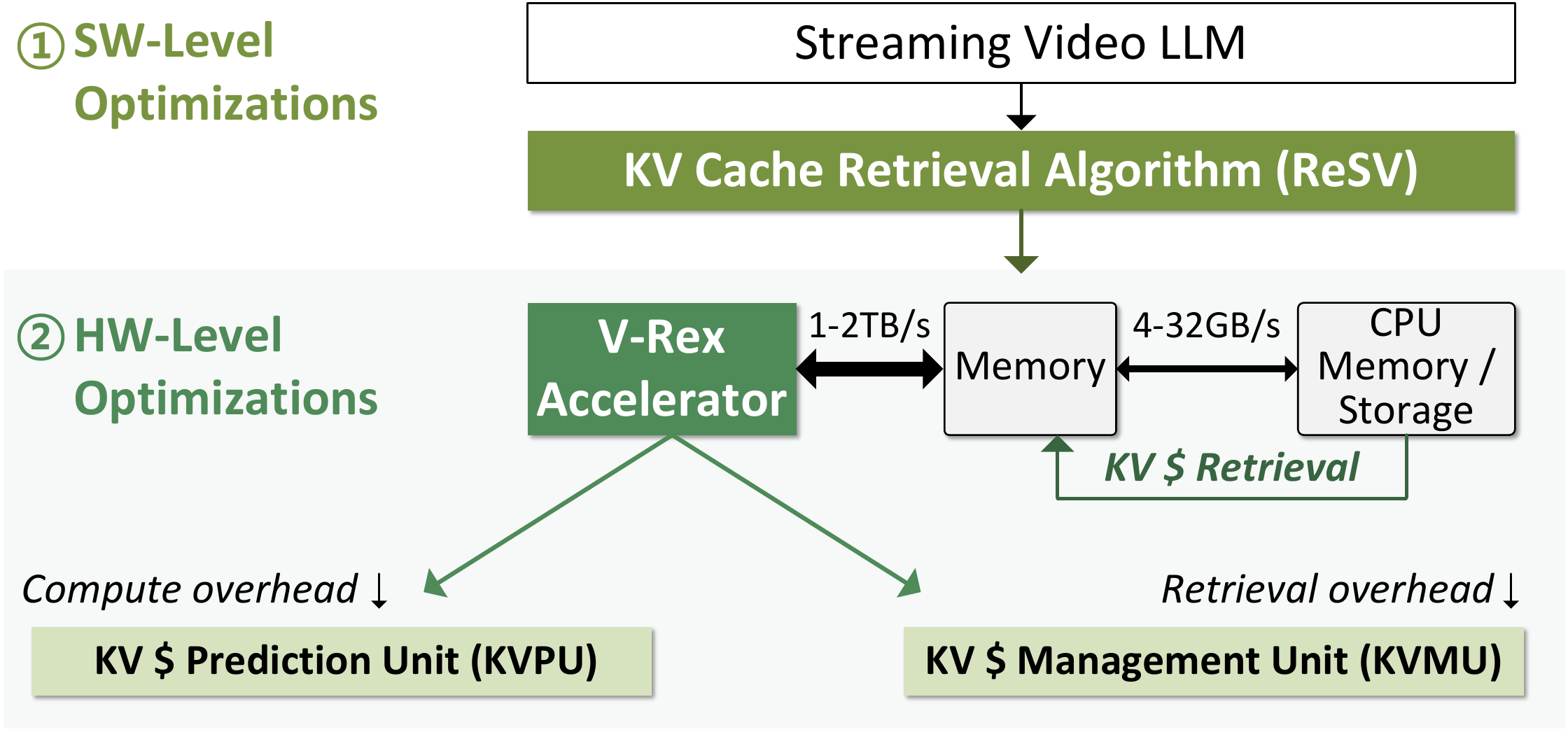}
\caption{Overview of \sysname Accelerator}
\label{f1}
\end{figure}

We present \sysname, the first streaming video LLM accelerator designed to address the large memory and computational requirements of the KV cache. It embodies this software-hardware co-design principle through the tightly integrated innovations, as shown in Figure~\ref{f1}. At the software level, we propose \algoname a training-free KV cache retrieval algorithm that intelligently perceives and exploits the unique characteristics of video data. It significantly reduces the number of fetched tokens for the iterative prefill stage. ReSV's hash-bit key clustering provides a computationally lightweight mechanism to identify and group tokens with high spatial-temporal similarity, drastically reducing redundancy without expensive computation. Building on this, its weighted cumulative sum (WiCSum) thresholding acts as an adaptive mechanism, dynamically selecting the most critical tokens on a fine-grained, layer-wise, and head-wise basis, moving far beyond the rigid constraints of fixed top-k. At the hardware level, we introduce the dynamic KV cache retrieval engine (DRE), a compact accelerator that serves as the essential enabler for \algoname. The KV cache prediction unit (KVPU) of DRE is specifically designed to execute the fine-grained, data-dependent, and conditional operations of \algoname, such as bit-level clustering and thresholding with early-exit sorting, that would cause severe slowdown and underutilization on a GPU. Additionally, the KV cache management unit (KVMU) of DRE complements this by managing PCIe bandwidth efficiently and reducing overall data movement during retrieval. By offloading these irregular tasks to a specialized unit, \sysname allows the main LLM engine to operate at peak efficiency.

The key contributions of this work are as follows:

\begin{itemize}
\item We propose \sysname, the first software-hardware co-designed accelerator that fundamentally addresses the large memory and computational bottleneck of the KV cache resulting from the iterative prefill stage in streaming video LLMs.

\item We introduce \algoname, a novel, training-free retrieval algorithm tailored for streaming video LLMs that leverages spatial-temporal similarity cache clustering and dynamic cache selection that reduces KV cache traffic with negligible accuracy loss.   

\item We developed the DRE, an efficient hardware unit that accelerates ReSV's irregular operations, making intelligent, fine-grained retrieval practical on resource-constrained platforms, consuming only 2.0\% of total chip area. It can be integrated with any existing GPUs, NPUs, and LLM accelerators with its high adaptability.

\item We demonstrate through comprehensive evaluation that \sysname enables real-time inference (3.9–8.3 FPS) on edge devices, achieving up to 19.7$\times$ speedup and 18.5$\times$ energy savings over a state-of-the-art GPU baseline.
\end{itemize} 
\section{Background and Motivations}
\label{sec_background}

\begin{figure}[t]
\centering
\includegraphics[width=0.49\textwidth]{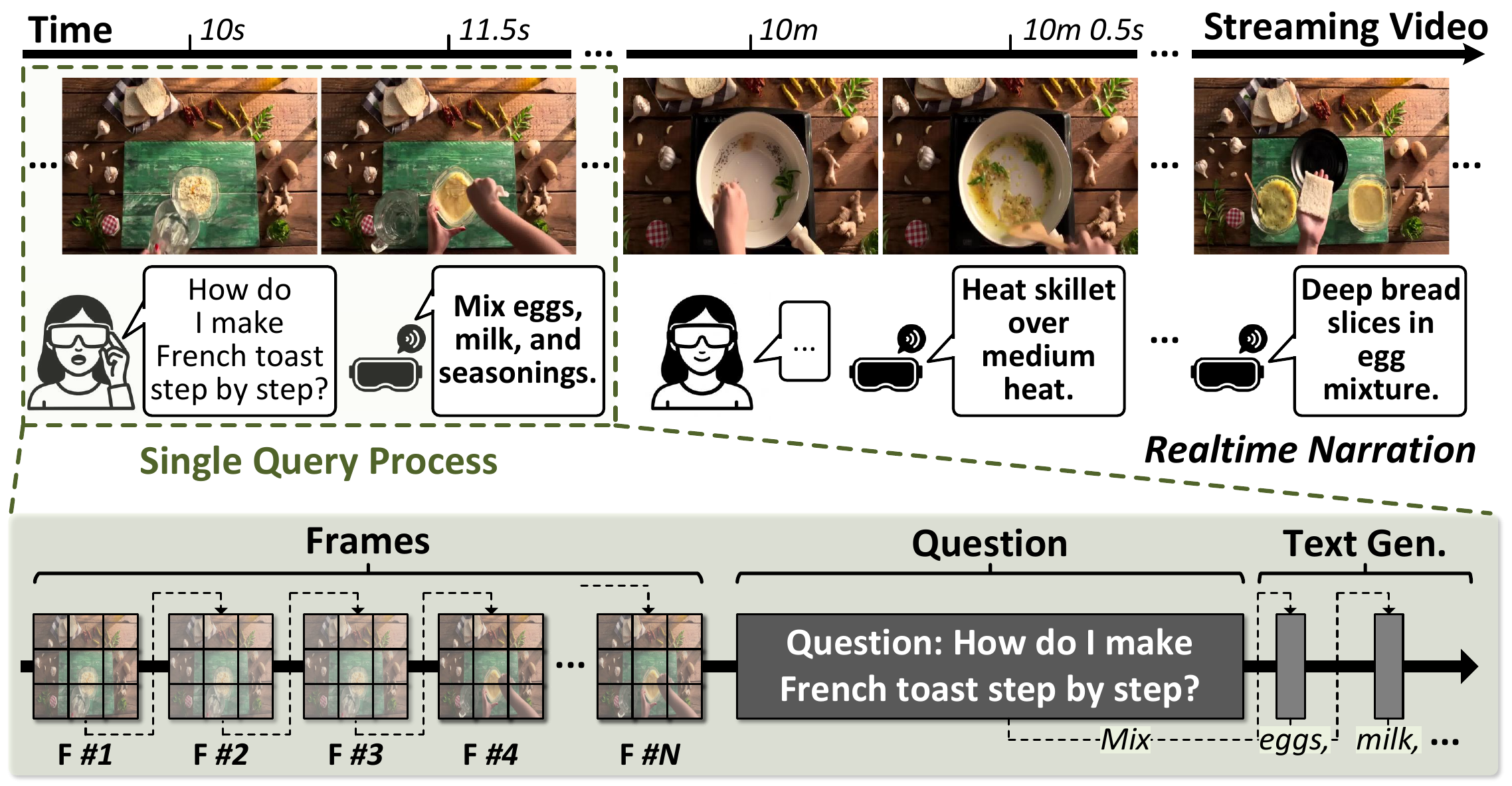}
\caption{Overview of Streaming Video LLM}
\label{f2}
\end{figure}
 
\subsection{Streaming Video LLM Architecture and Workflow}

Figure~\ref{f2} presents an overview of the streaming video LLM. Unlike offline models, it processes real-time streaming video input and generates narration in direct response to user queries. Users may issue a series of queries, including follow-ups that depend on both previous interactions and the evolving video context. Consequently, information from earlier video segments is vital for producing context-aware responses to future queries. This operational need underscores the importance of advanced KV cache management algorithms that preserve and utilize prior visual context, rather than relying on conventional optimization methods (i.e., pruning, merging, and quantization)~\cite{tao2025plug, li2025commvq, liu2024kivi, kim2025oaken,10.5555/3737916.3742359, kim2025infinipot, tu2024vl, wu2024videollm} that may discard information essential for subsequent interactions.

\begin{figure}[t]
\centering
\includegraphics[width=0.49\textwidth]{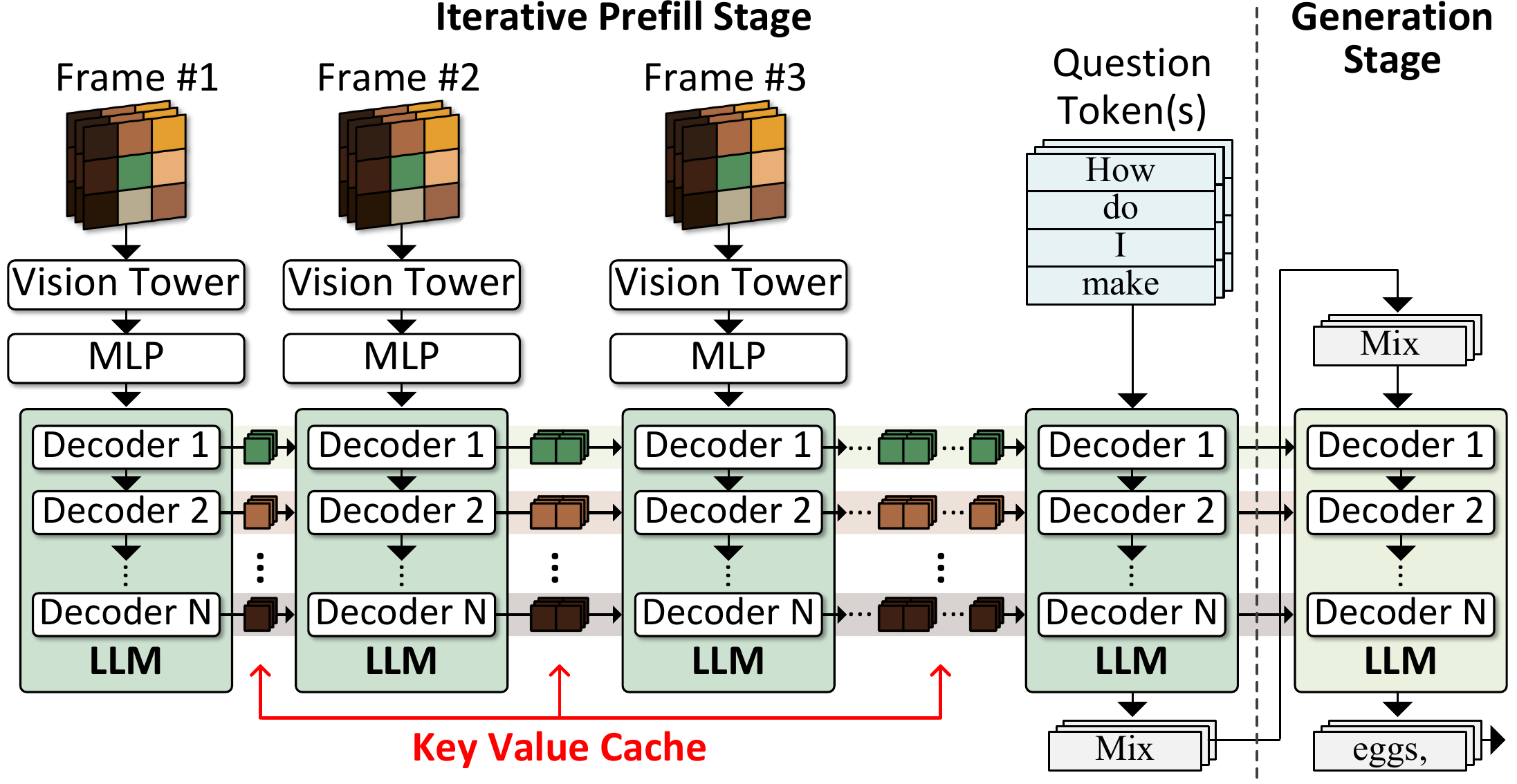}
\caption{Model Architecture of Streaming Video LLM}
\label{f2b}
\end{figure}

Figure~\ref{f2b} shows the model architecture of streaming video LLM. A streaming video LLM consists of three core modules: a vision tower, an MLP projector, and an LLM. The vision tower (e.g., CLIP~\cite{pmlr-v139-radford21a}, SigLIP~\cite{Zhai_2023_ICCV}) transforms each video frame into numerical embeddings that capture spatial and temporal features. The MLP projector adapts the dimension of these embeddings, enabling seamless integration with the LLM input space. The LLM processes visual information and user queries, generating output tokens. For the LLM, models such as Llama-3~\cite{2024arXiv240721783G} and Qwen3~\cite{yang2025qwen3technicalreport} are often used.

\hypertarget{target:A_2_1}{}\textcolor{black}{The streaming video LLM first performs \textbf{iterative prefill stage} that sequentially processes video inputs and question tokens,} a distinctive mechanism unique to handling real-time video streams. \textcolor{black}{Since sampled frames in a real-time video stream arrive sequentially and cannot be batched together,} each frame is processed individually through a repeated prefill computation of LLM. Each prefill stage \textcolor{black}{attends previous KV cache for the self-attention computation and} generates KV cache entries that are incrementally accumulated. This KV cache grows \hypertarget{target:A_1_1}{}\textcolor{black}{linearly over time}, following an $O(N^{2}T)$ complexity, where $N^2$ represents the spatial resolution and $T$ denotes temporal duration. Notably, this cache facilitates the processing of future frames and is crucial for generating accurate responses to user questions, as queries may reference visual information spanning multiple frames. When the user inputs a query, the user’s question is tokenized and processed solely through the LLM. \textcolor{black}{In the generation stage,} it generates output based on both the accumulated frame KV caches and the question tokens, thereby maintaining contextual coherence.

\subsection{Principles of KV Cache Retrieval}

Figure~\ref{f3} (a) shows the overhead of the KV cache of VideoLLM-Online~\cite{chen2024videollm} when using Llama-3 8B as the backbone model. The KV cache increases with video duration and exceeds GPU memory capacity within minutes, making edge deployment impractical. Prior research attempts to reduce KV cache size through pruning and merging, but it falls short for streaming video LLMs in multi-turn settings. Discarding segments of the cache results in inaccurate responses to sequential user queries. In contrast, KV cache retrieval preserves all prior information and enables selective computation, thereby reducing memory requirements while preserving model accuracy. This is achieved through a three-stage process during inference. (1) Offloading: the entire KV cache is first transferred to CPU memory or storage. (2) Selection: only relevant tokens are selected for the query. (3) Pre-fetching: these selected KV entries are retrieved to the GPU memory in advance for attention computation. This design ensures three essential outcomes: 1) It upholds contextual integrity across multi-turn queries, 2) minimizes the GPU memory requirements, and 3) reduces computation by limiting processing to the most relevant subset of the cache per query. Thus, KV cache retrieval offers both scalability and coherence for real-time streaming video LLMs.

\section{Challenges of KV Cache Retrievals}

\subsection{Why KV Retrievals Fall Short in Streaming Video LLMs}

\hypertarget{target:A_2_2}{}Applying existing KV cache retrieval techniques to streaming video LLMs poses fundamental limitations that have not been addressed in prior works. For instance, InfiniGen~\cite{lee2024infinigen} is a representative algorithm that effectively hides retrieval latency during the LLM’s generation stage. However, in real-world streaming video LLM scenarios, this advantage has minimal impact because such systems are dominated by an iterative prefill stage\textcolor{black}{, which utilizes KV caches,} driven by continuous incoming frames and multi-turn queries. InfiniGen and other similar approaches operate exclusively during generation and thus do not address the predominant bottleneck during prefill, where the bulk of KV cache retrieval and generation occurs.

\begin{figure}[t]
\centering
\includegraphics[width=0.49\textwidth]{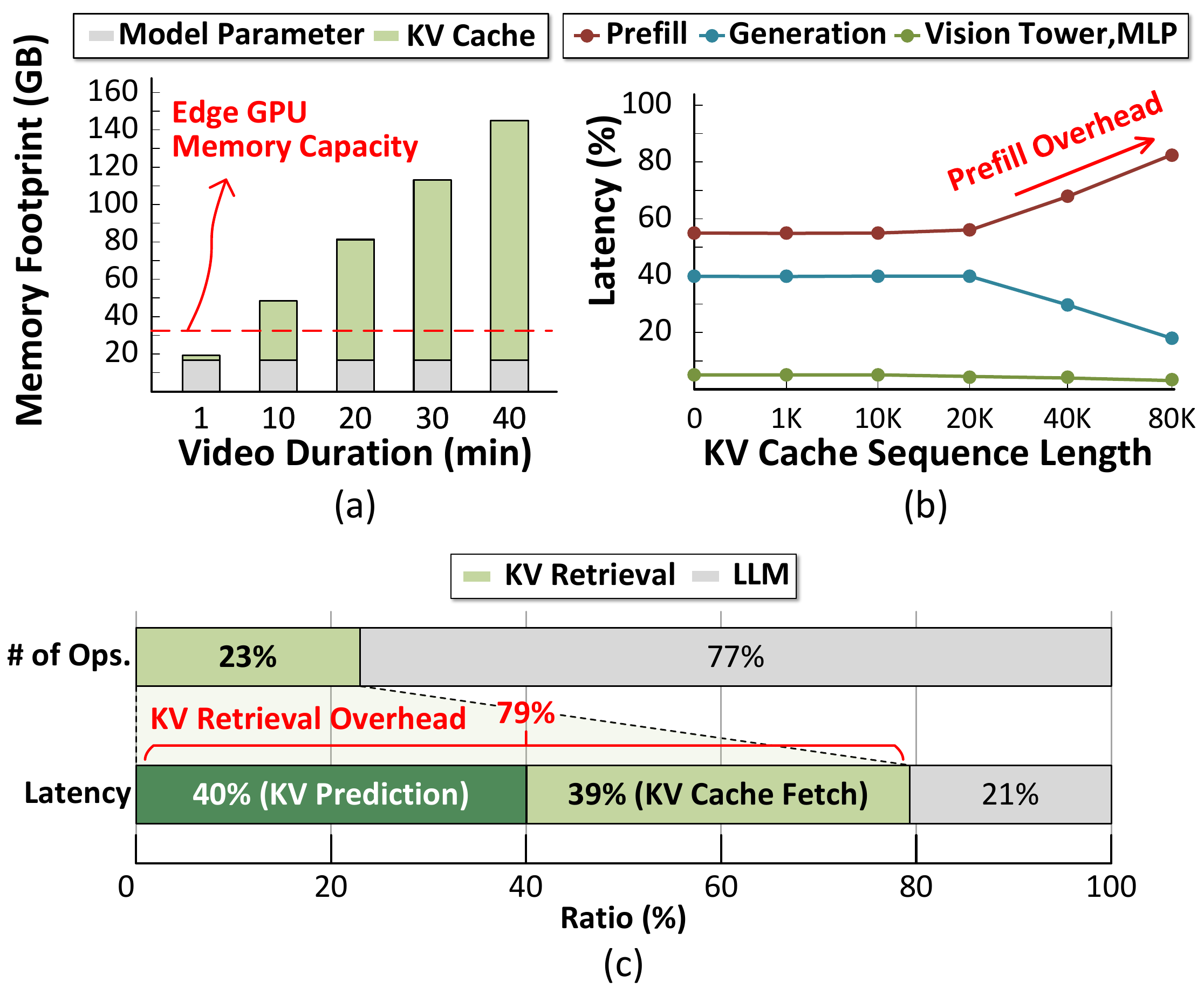}
\caption{(a) Memory Footprint of Streaming Video LLM under a 10FPS setting at batch 4. \textcolor{black}{(b) End-to-end Latency Breakdown of Streaming Video LLM.} (c) KV Retrieval Latency Overhead of SOTA Retrieval Method~\cite{lee2024infinigen} in Prefill Stage \textcolor{black}{at 40K KV Cache Sequence Length}.}
\label{f3}
\end{figure}

We \textcolor{black}{analyzed the breakdown of end-to-end latency of streaming video LLM using InfiniGen on an NVIDIA A100 GPU by modeling the average working scenario on the COIN benchmark (i.e., 26 frames, 25 question tokens, and 39 answer tokens), assuming a specific length of the KV cache sequence has been pre-computed and is actively maintained}, as shown in Figure~\ref{f3} (b). The results reveal that as video duration increases, the number of accumulated KV cache tokens grows rapidly, making prefill the largest contributor. \textcolor{black}{At 80K cache sequence length, 83\% of the latency is taken by the prefill stage and 74\% of this prefill latency is taken by the KV cache retrieval, confirming it is the true bottleneck.} Since the prior retrieval method only optimizes the generation stage, it fundamentally fails to tackle the most critical memory and performance bottlenecks in streaming video LLMs. Without directly addressing KV cache management during frequent prefill, it is not possible to achieve practical gains in memory efficiency or system responsiveness under streaming workloads.

\subsection{Limitations of Adapting Retrieval Algorithms}
Adapting GPU-oriented retrieval algorithms (FlexGen~\cite{sheng2023flexgen}, InfiniGen, ReKV~\cite{di2025streaming}) to streaming video LLM prefill stages causes significant inefficiency due to KV prediction computation and CPU-GPU data transfer overhead. The computation overhead for KV prediction increases as the KV cache sequence increases. In addition, in streaming scenarios, the \textit{Query} matrix consists of multiple tokens, each requiring different KV cache entries, necessitating larger token budgets than those for text generation. To empirically illustrate these issues, we measure the latency breakdown of streaming video LLM at 40K KV cache sequence length when InfiniGen is adopted for the prefill stage \hypertarget{target:A_4_1}{}\textcolor{black}{with token budget of 10K} on an NVIDIA A100 GPU, as depicted in Figure~\ref{f3} (c). The KV cache retrieval includes both KV prediction computation and memory transfer latencies. Results show that the KV cache retrieval computation only accounts for 23\%. However, it accounts for 85\% of the total latency, where 40\% is attributed to the KV prediction computation and 39\% to the KV cache fetch from CPU memory. We additionally confirmed this issue with other SOTA retrieval methods (i.e., FlexGen and ReKV), demonstrating a similar trend. Furthermore, this issue becomes more severe as the token sequence length increases, causing larger KV prediction computation and data retrieval. These results highlight that existing GPU-oriented algorithms cannot efficiently handle prefill-heavy streaming workloads. Addressing this bottleneck requires fundamentally new approaches.

\begin{figure}[t]
\centering
\includegraphics[width=0.49\textwidth]{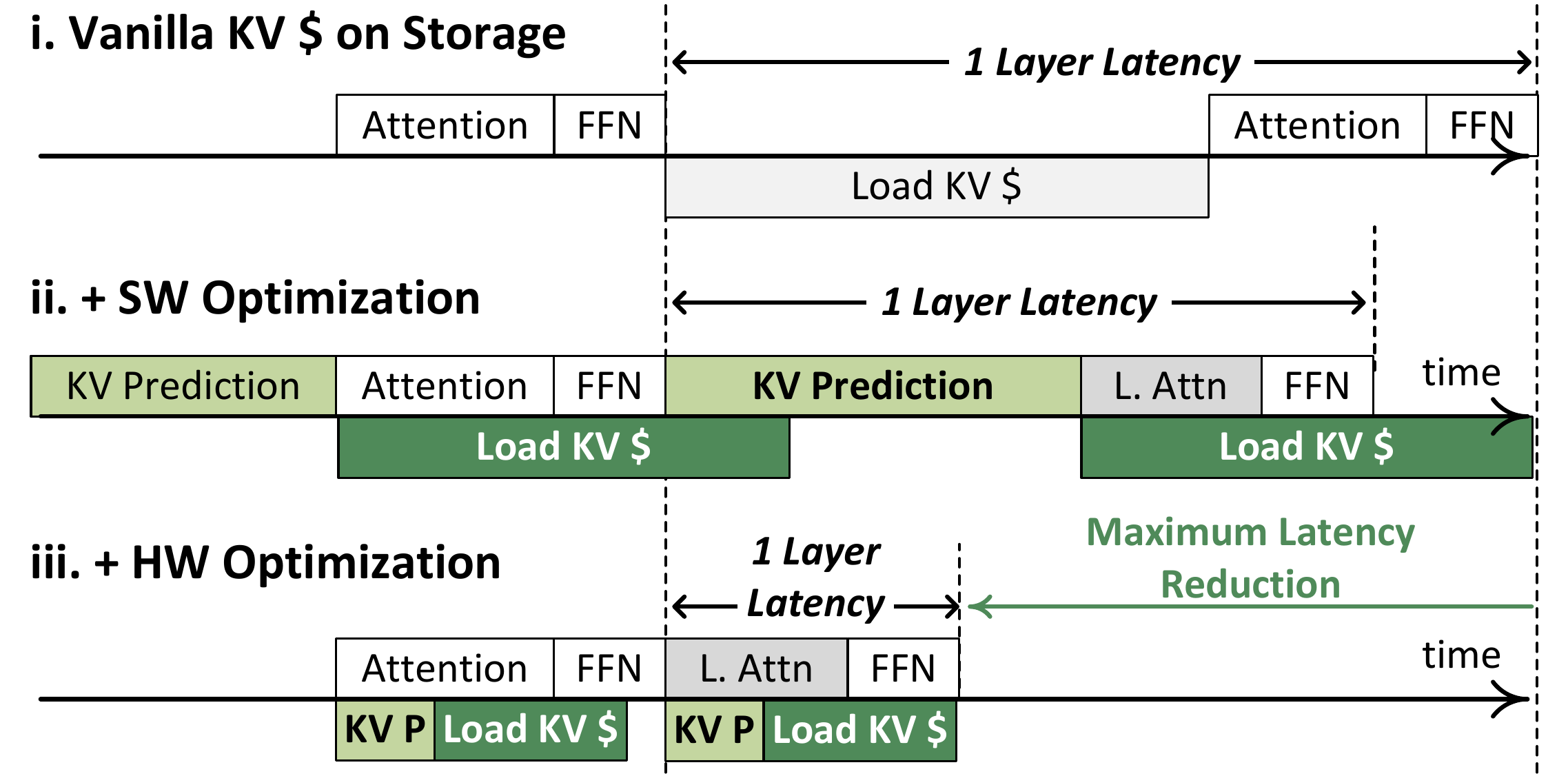}
\caption{\sysnames Software-Hardware Co-design Optimization}
\label{f4}
\end{figure}

\subsection{Inflexibility of Fixed Top-K Selection}

Many GPU-oriented algorithms, including InfiniGen and ReKV, favor top-k selection in KV cache management to take advantage of the predictable resource allocation and efficient parallel processing inherent to GPU architecture. However, this static approach imposes fundamental limitations for streaming video LLMs. Crucially, the score matrices that determine token importance vary widely across different transformer layers and attention heads, \hypertarget{target:A_7_1}{}\textcolor{black}{reflecting that diverse features are captured throughout them. Consequently, a different set of tokens is selected as important by each unique layer and head.} Prior studies have shown that fixed top-k selection frequently results in redundant tokens or loss of relevant tokens, since the optimal K shifts by layer and head~\cite{tu2024vl, yang2024pyramidinfer, ge2023model}.

These inefficiencies are exacerbated in streaming edge environments, where memory budgets are limited and strict latency constraints apply. In such contexts, over-provisioning KV cache due to inflexible top-k policies leads to avoidable resource overhead and longer response times, undermining system scalability and energy efficiency. Additionally, the nature of streaming video LLMs requires the video data to be streamed, and the sequence length increases in real-time, necessitating the adaptive adjustment of the number of selected tokens to ensure efficiency and accuracy. To this end, \sysname is explicitly designed to address these challenges, providing fine-grained, importance-driven dynamic selection that reduces KV cache size and retrieval cost for practical, real-time inference even on resource-constrained edge platforms.

\section{V-Rex: Unified SW-HW Co-Design Strategy}
To address the challenges of streaming video LLMs, we propose \sysname, a software-hardware co-designed solution. Figure~\ref{f4} illustrates how each component of our optimization stack reduces decoder layer latency. At the software level, \sysname implements \algoname, an enhanced KV cache retrieval algorithm that efficiently selects and fetches only the most relevant entries from CPU memory or storage, where full caches are offloaded. It improves upon prior methods by using hash-bit key clustering and WiCSum thresholding. Leveraging the high temporal and spatial similarity in video frames, the algorithm achieves lightweight computation and efficient KV selection. At the hardware level, \sysname integrates compact units to accelerate these operations and minimize retrieval overhead. It decouples these operations from the main LLM computation pipeline, enabling latency hiding and end-to-end optimization.

\begin{figure}[t]
\centering
\includegraphics[width=0.49\textwidth]{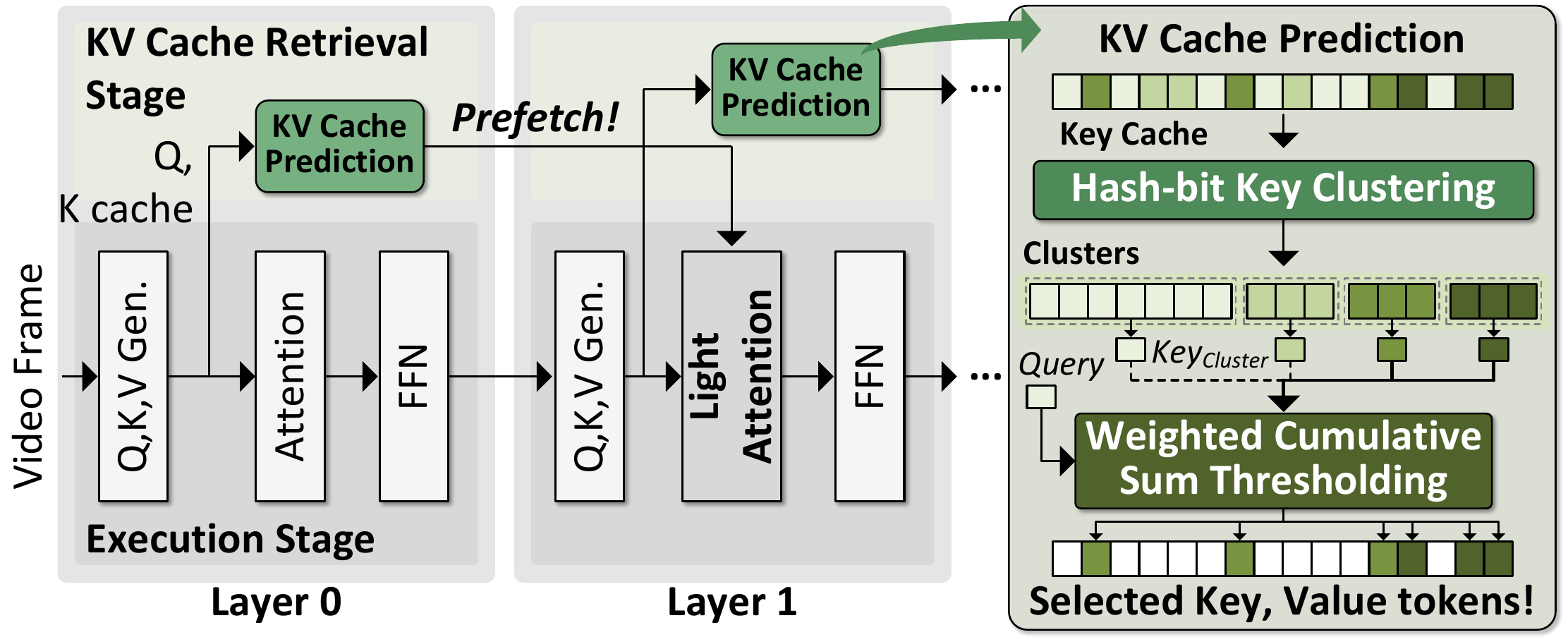}
\caption{Overview of \algoname Algorithm}
\label{f5}
\end{figure}

\subsection{\algoname: Efficient and Accurate KV Cache Retrieval}

ReSV is a training-free algorithm designed to optimize KV cache retrieval during the iterative prefill stage of streaming video LLMs. As shown in Figure~\ref{f5}, it consists of two main stages: KV retrieval and execution. In the retrieval stage, KV prediction is performed on-the-fly immediately after QKV generation to capture the dynamic nature of the cache. Selected KV tokens are prefetched for the next decoder layer, hiding fetch latency during execution. KV prediction involves two steps. First, hash-bit key clustering groups similar tokens using lightweight bitwise operations, generating representative keys (\textit{Key}$_{cluster}$) by averaging within each cluster. This reduces computation by limiting attention to representative keys. Second, WiCSum thresholding dynamically selects the most important \textit{Key}$_{cluster}$, adapting to varying data distributions across layers and attention heads, unlike fixed top-k methods. In the execution stage, the model performs light attention using only the selected clusters, significantly reducing memory and compute by fetching only essential KV entries.

\subsection{Hash-bit Key Clustering for Fast Similarity Grouping}

\begin{figure}[t]
\centering
\includegraphics[width=0.49\textwidth]{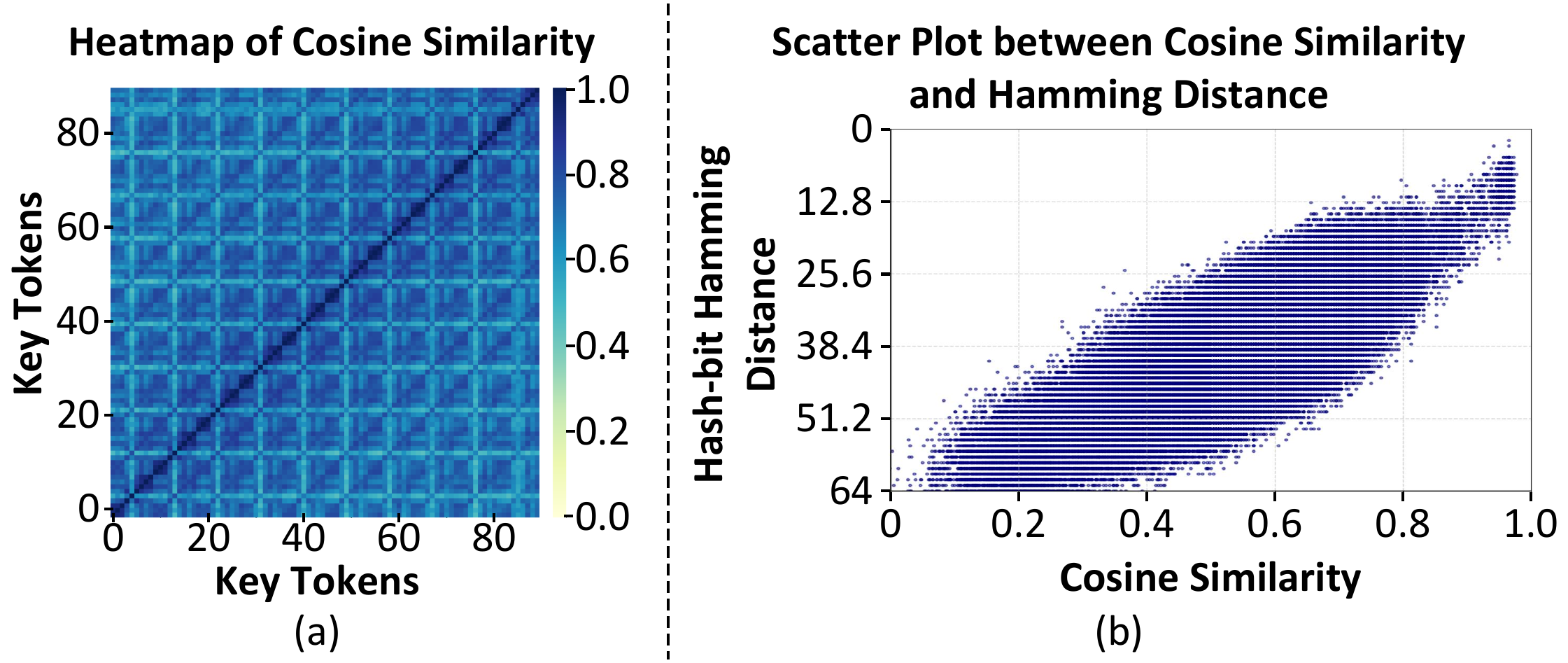}
\caption{(a) Heatmap of Cosine Similarity of Key Tokens Between Adjacent Frames (b) Scatter Plot Between Cosine Similarity and Hash-bit Hamming Distance. Measured on 3$^{rd}$ Layer's Key using COIN Dataset.}
\label{f6}
\end{figure}

The rationale for hash-bit key clustering lies in the high similarity among tokens in adjacent frames, as shown in Figure~\ref{f6} (a). Leveraging this property, the method performs spatial-temporal clustering of key caches to efficiently reduce redundancy across frames. Unlike merging methods that replace multiple tokens with a single representation which requires higher precision, this approach preserves original token values for the execution stage. Thus, it avoids expensive operations like high-dimensional cosine similarity by using ultra-low-dimensional representations ($\leq$ 0.5\% of the original dimension) and lightweight hash-bit hamming distance computations. Figure~\ref{f6} (b) proves that our hash-bit hamming distance can effectively follow the trend of cosine similarity, having a 0.8 correlation value, which is enough to perform clustering.

\begin{figure}[t]
\centering
\includegraphics[width=0.49\textwidth]{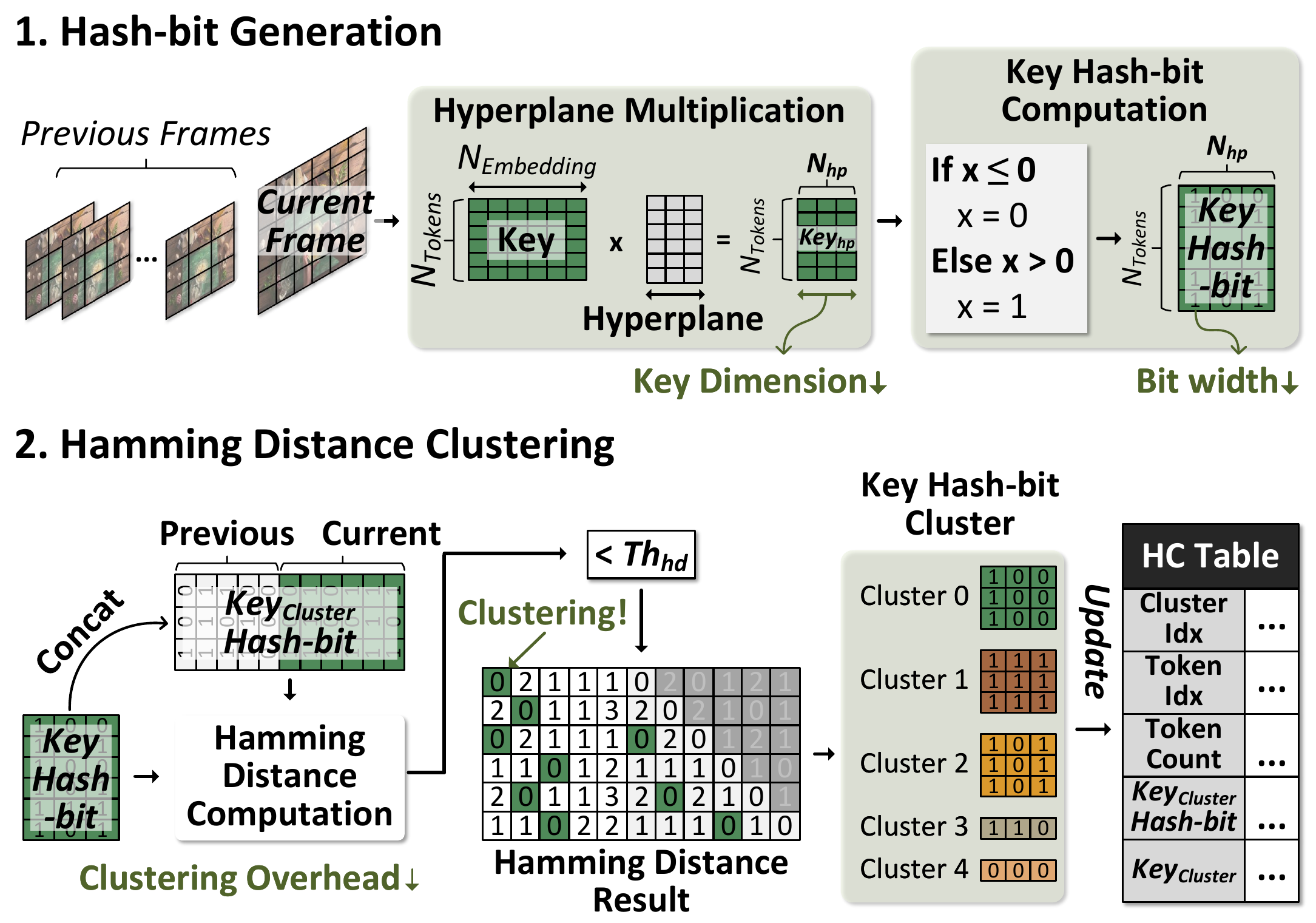}
\caption{Dataflow of Hash-bit Key Clustering}
\label{f7}
\end{figure}

The clustering process consists of two main steps: hash-bit generation and hamming distance clustering, as shown in Figure~\ref{f7}. In the hash-bit generation step, computation is performed each time a new frame arrives. The key matrix, obtained after applying the rotary position embedding operation to the current frame, undergoes dimensionality reduction in two directions to significantly reduce the overhead of the following hamming distance clustering. A set of $N_{hp}$ random hyperplanes is generated, and the key matrix is multiplied by these hyperplanes, producing a reduced-dimension matrix $Key_{hp}$ with $N_{hp}$ embedding dimensions. Each element of $Key_{hp}$ is then binarized: values less than or equal to zero are set to 0, and values greater than zero are set to 1, converting each element into a single bit to form the \textit{Key hash-bit}.

Next, hamming distance clustering is performed. It involves calculating hamming distance between the newly generated current \textit{Key} \textit{hash-bit} and the combined \textit{Key}$_{cluster}$ \textit{hash-bit}, which includes previous and current \textit{Key hash-bits}. The hamming distance is computed by performing a bit-wise XOR operation between tokens and counting the number of differing bits. Tokens with distances below a hyperparameter-defined threshold ($Th_{hp}$) are clustered. The final clustering results are stored in a hash cluster (HC) table containing the cluster index, token index, \textit{Key}$_{cluster}$, \textit{Key}$_{cluster}$ \textit{hash-bit}, and token count. This method enables clustering with very low computational overhead that typically grows with token count while maintaining high clustering accuracy, making it well-suited for efficient KV cache selection in streaming video LLMs.

\begin{figure}[t]
\centering
\includegraphics[width=0.49\textwidth]{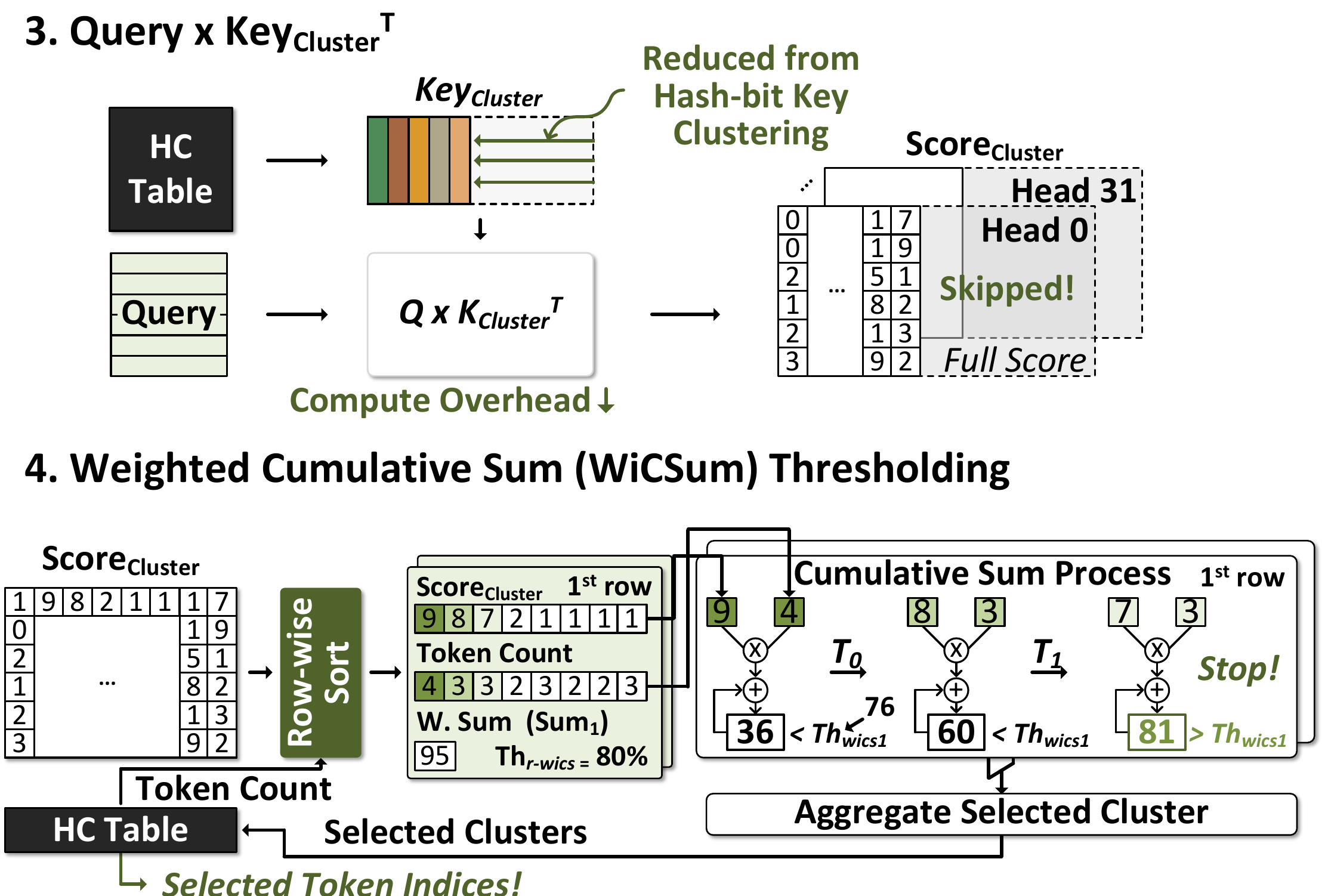}
\caption{Dataflow of Weighted Cumulative Sum Thresholding}
\label{f8}
\end{figure}

\begin{figure*}[t]
\centering
\includegraphics[width=0.99\textwidth]{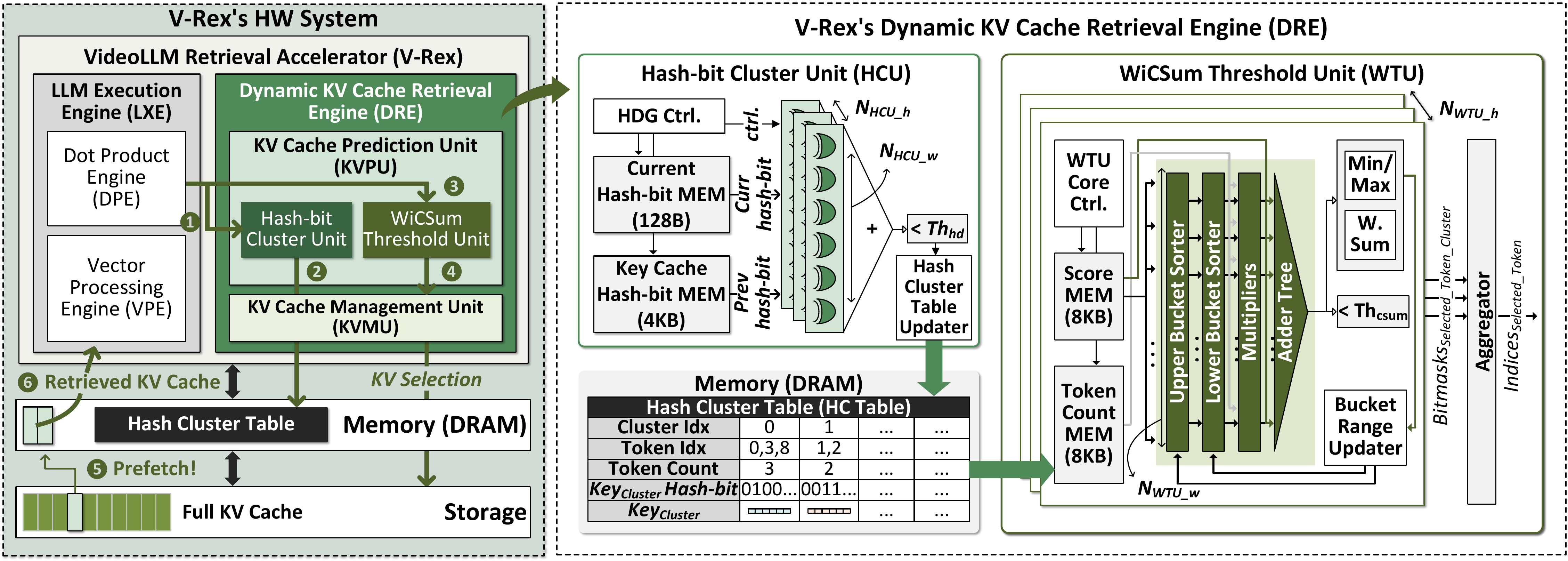}
\caption{\textcolor{black}{Overall Architecture of \sysname}}
\label{f9}
\end{figure*}

\subsection{Dynamic Token Selection via WiCSum Thresholding}
WiCSum thresholding is a dynamic selection algorithm developed to address the diverse score distributions that occur across different layers and attention heads. Unlike static top-k methods that select a fixed number of tokens regardless of their importance, WiCSum thresholding dynamically determines the number of tokens to select for each layer and head. This dynamic approach enables more accurate and efficient KV cache retrieval, minimizing unnecessary memory and computational overhead, thereby supporting low latency and high efficiency.

Figure~\ref{f8} shows the overall dataflow, composed of two main steps: \textit{Query} × \textit{Key}$_{cluster}^T$ computation and threshold checking. In the first step, the algorithm computes the matrix multiplication between the current query vectors and the representative \textit{Key}$_{cluster}$ generated by the previous hash-bit key clustering stage. Because this computation uses only the representative \textit{Key}$_{cluster}$ values rather than the entire key cache, it significantly reduces the computational overhead. The result of this operation is the \textit{Score}$_{cluster}$ matrix, which reflects the relevance of each \textit{Key}$_{cluster}$ to the current query.

In the threshold checking step, important elements in the \textit{Score}$_{cluster}$ matrix are selected. For each row \textcolor{black}{\textit{i}} in the matrix, it calculates a weighted sum ($Sum_i$) by multiplying each score by its corresponding token count and summing the results, \textcolor{black} { as shown in Equation 1.} This weighted sum is then used to compute a threshold ($Th_{wics}$) by multiplying it by a predefined ratio hyperparameter ($Th_{r-wics}$), \textcolor{black}{as shown in Equation 2.} Then, each row of \textit{Score}$_{cluster}$ is sorted in descending order, \textcolor
{black}{where $\sigma$ is the sorting permutation.} Starting from the highest \textit{Score}$_{cluster}$ value, the weighted sum with the token count is accumulated until the \textcolor{black}{minimum \textit{t}, when $Acc_i(t)$} exceeds the threshold $Th_{wics_i}$, \textcolor{black}{as shown in Equation 3.} The indices of the clusters selected up to this point are aggregated across all rows, and these selected cluster indices are then mapped back to the original token indices using the HC table to produce the final set of selected tokens.
\hypertarget{target:A_8_1}{}
\begin{equation}
\textcolor{black}{Sum_i = \sum_{j=0}^{cluster} Score_{cluster_{i,j}} \cdot TC_j}
\label{reordering}
\end{equation}
\begin{equation}
\textcolor{black}{Th_{wics_i} = Sum_i \cdot Th_{r-wics}}
\label{reordering}
\end{equation}
\begin{equation}
\textcolor{black}{Acc_i(t) = \sum_{j=0}^{t} Score_{Cluster_{i,\sigma(j)}} \cdot TC_{\sigma(j)}, \space Acc_i(t) > Th_{wics_i}}
\label{reordering}
\end{equation}

\section{\sysnames Hardware Architecture}
\hypertarget{target:A_9_1}{}The \algoname effectively reduces the number of required tokens. Nevertheless, the core operations introduced by \algoname present inefficiencies on GPUs. These inefficiencies arise from \textcolor{black}{1) conditional and data-dependent computation of \algoname's clustering and thresholding, which makes it difficult to fully exploit parallelism, and 2) irregular and sparse KV cache fetching from SSD and CPU memory, which causes under-utilization of PCIe bandwidth, both resulting in increased latency.} To address these challenges, we introduce \sysname with a low-latency, compact KV cache retrieval engine specifically designed to support the unique computational patterns of \algoname \textcolor{black}{and optimize the memory-intensive KV fetching by efficiently handling the irregular memory access patterns}. Additionally, it can be easily integrated with existing hardware, including GPUs and NPUs, for high adaptability.

\subsection{Architecture Overview}
Figure~\ref{f9} illustrates \sysnames hardware system, which consists of three primary components: the \sysname accelerator, off-chip memory, and storage or CPU memory for the full KV cache. The \sysname accelerator, which comprises the LLM execution engine (LXE) and DRE, is responsible for the primary computational tasks required by streaming video LLMs. \textcolor{black}{The execution flow proceeds as follows: \circled{1} LXE generates hash-bits for current frame keys, \circled{2} hash-bit cluster unit (HCU) performs hamming distance clustering and updates HC table, \circled{3}  LXE computes \textit{Q} × \textit{K}$_{cluster}^T$ then send \textit{Score}$_{cluster}$ to WiCSum threshold unit (WTU), \circled{4} WTU executes WiCSum thresholding with early-exit sorting, determining which token entries to retrieve, \circled{5} KVMU prefetches selected KV entries from storage, and \circled{6} retrieved KV tokens are used in attention.}

\textbf{LLM Execution Engine.} LXE processes the primary LLM operations and two computations from \algoname. The hash-bit generation and \textit{Query} × \textit{Key}$_{cluster}^T$ computation of \algoname are processed in LXE, as it involves mainly matrix multiplications and vector computations. The LXE is based on the core architecture of the LPU~\cite{10591630}, which integrates a dot product engine (DPE) for high-throughput matrix multiplication and a vector processing engine (VPE) for efficient vector operations, both with BF16 precision. DPE is composed of $N_{DPE-h}$ MAC trees, receiving $N_{DPE-w}$ inputs. The VPE is composed of $N_{VPE-h}$ vector units and accepting $N_{VPE-w}$ inputs. 

\subsection{Dynamic KV Cache Retrieval Engine (DRE)}
The DRE consists of a 
KVPU and KVMU, which are responsible for the computations and memory management required during dynamic KV cache retrieval. The KVPU integrates both HCU and WTU, which together accelerate the most latency-critical operations in KV cache retrieval. By offloading these tasks from the main compute pipeline, \sysname significantly reduces computational and data fetching bottlenecks.

\textbf{Hash-bit Cluster Unit.} At the core of the KVPU, the HCU is responsible for executing the hash-bit key clustering process. The HCU is a compact computing module, consisting of three main components: a current hash-bit memory, a key cache hash-bit memory, and $N_{HCU-h}$ parallel XOR accumulators, each capable of processing $N_{HCU-w}$ inputs. The HCU receives the key hash-bit vectors from the LXE and stores them in the current hash-bit memory. Simultaneously, it retrieves key cache hash-bit clusters from the HC table and stores them in the key cache hash-bit memory. Both of these are structured as bit matrices to enable efficient parallel operations.

To perform clustering, the HCU initiates the computation of hamming distances between the current hash-bit vectors and the stored key cache hash-bit clusters. This process utilizes XOR accumulators to identify differences between corresponding bits across the matrices. The accumulators then sum the number of differing bits to calculate the hamming distance for each comparison. By comparing the computed hamming distances with the predefined threshold $Th_{hd}$, the HCU efficiently groups tokens with similar hash-bit patterns into clusters. Then, the clustering results are stored in the HC table. This hardware-accelerated approach enables rapid and energy-efficient clustering using bitwise operators, supporting the low-area requirements for edge deployment.

\begin{figure}[t]
\centering
\includegraphics[width=0.49\textwidth]{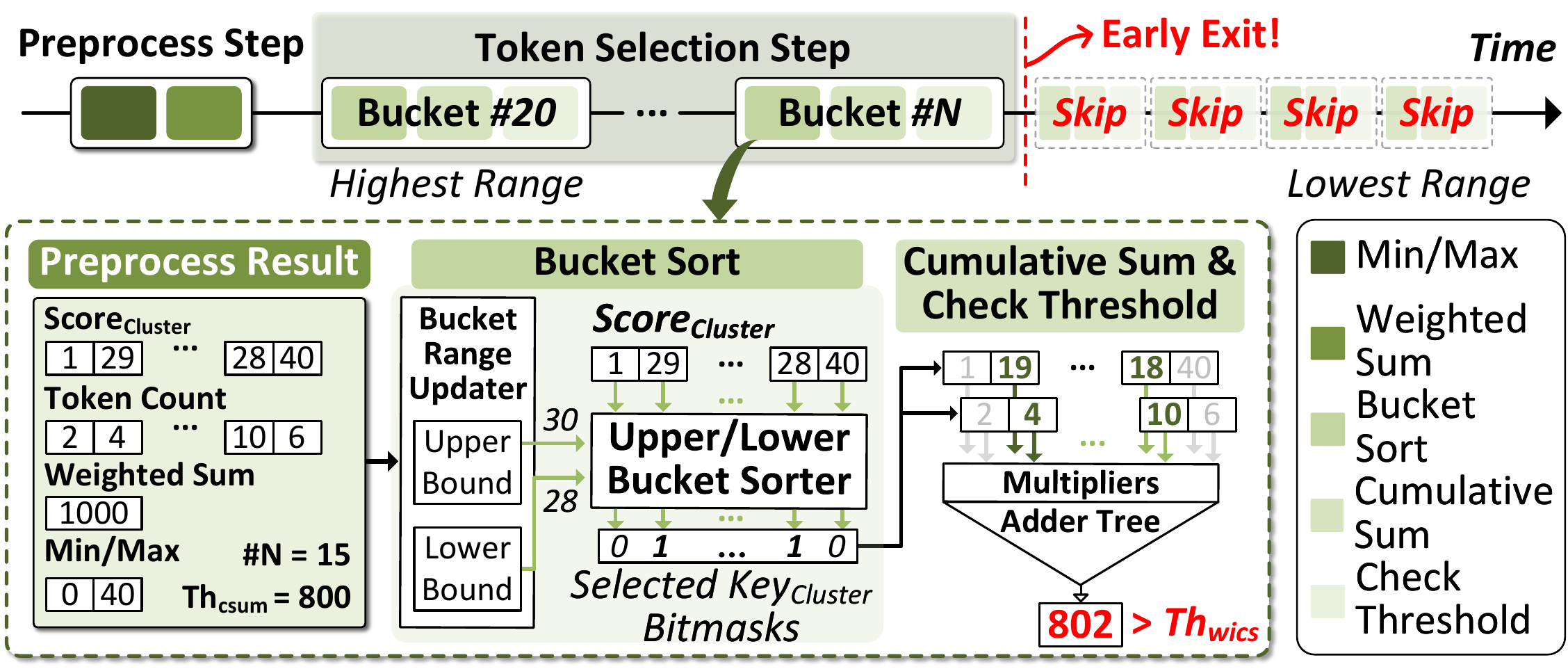}
\caption{Dataflow of Early Exit Sorting}
\label{f10}
\end{figure}

\textbf{WiCSum Threshold Unit.} The WTU accelerates the WiCSum threshold check, enabling low-latency selection computation. It consists of multiple parallel WTU cores, each equipped with score memory, token count memory, and a dedicated computing unit for thresholding. Each core includes upper and lower bucket sorters, multipliers, an adder tree, and a bucket range updater. The most computationally intensive operations, sorting and threshold checking, are efficiently handled by the WTU’s dataflow, which utilizes early exit sorting. It combines two operations in a fine-grained pipeline and terminates sorting in the middle when it exceeds the threshold, as shown in Figure~\ref{f10}. This process is divided into two main steps: the preprocess step and the token selection step. In the preprocess step, the WTU cores precompute values needed for the token selection step, such as the weighted sum of scores and token counts for each row, the min/max score values, and the threshold $Th_{wics}$. During the token selection step, the process begins with the bucket containing the highest score range. The WTU performs bucket sort, cumulative sum, and threshold checking in the pipeline. The bucket sort, which is highly parallelizable, uses the preprocessed information to determine the upper and lower bounds for each bucket, and the sorters generate bitmasks indicating which scores fall within the current range. The selected values are then multiplied and summed to compute the weighted sum, which is compared to $Th_{wics}$ to decide whether to exit or continue. This early exit mechanism is effective because a small number of large scores typically account for the majority of the weighted sum (average 16\% per row), allowing the threshold to be reached quickly by starting with the highest buckets.

\subsection{KV Cache Management Unit}
\begin{figure}[t]
\centering
\includegraphics[width=0.49\textwidth]{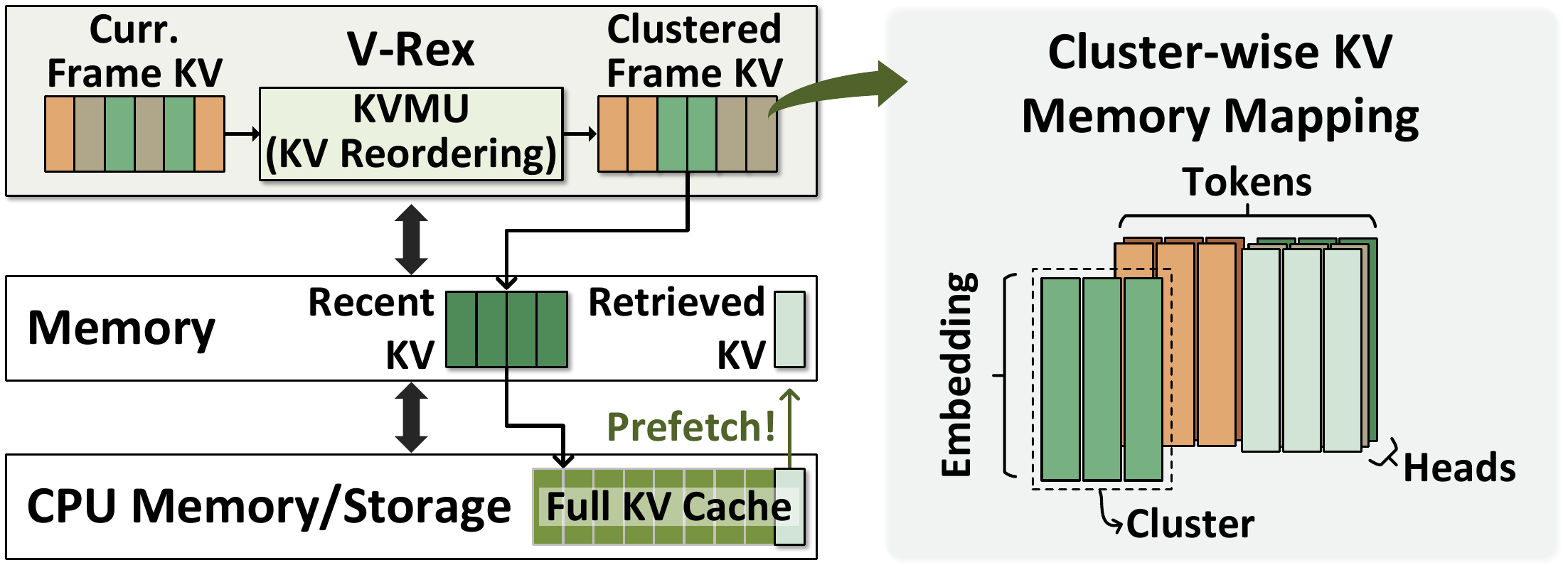}
\caption{Hierarchical Memory System and Cluster-wise Memory Mapping}
\label{f11}
\end{figure}

The KVMU manages the KV cache's memory for streaming video LLMs. It is responsible for two main functions: hierarchical KV cache memory management and hash cluster-based memory mapping. First, KVMU oversees a hierarchical memory system, as illustrated in Figure~\ref{f11}, to efficiently manage data movement between \sysnames memory, CPU memory, and storage. Recent KV cache entries are stored in \sysnames memory for fast access. When the total size of the KV cache in \sysnames memory exceeds a predefined maximum capacity, the oldest entries are offloaded to CPU memory or storage. These offloaded entries can be retrieved from CPU memory or storage and brought back into \sysnames memory when needed for computation. This hierarchical memory system ensures that both the most recent and retrieved KV cache entries are available for computation, while older or less critical data is kept off-chip to significantly reduce memory overhead. 

Second, KVMU implements an efficient memory mapping strategy based on hash clusters. To maximize PCIe bandwidth utilization, tokens that belong to the same hash cluster are grouped and stored at contiguous memory addresses. The clustering is carried out entirely within the recent KV cache, removing any need to access the CPU or storage for clustering with the offloaded cache. This arrangement enables more efficient use of memory bandwidth, as multiple tokens from the same cluster can be transferred together in a single operation. Each time new KV cache entries are generated for a frame, KVMU reorders and stores them in memory according to the latest clustering results. Because KVMU handles this reordering internally, the KV cache is stored in a streaming fashion, and any latency associated with reorganization is effectively hidden. \hypertarget{target:A_10_1}{}\textcolor{black}{Although this memory mapping is technically feasible on conventional GPUs, it is highly impractical because it requires fine-grained, online data reorganization. This process incurs substantial latency overhead that ultimately nullifies the benefits of the optimized layout, as it involves frequent per-layer computations and irregular, memory‑intensive scattering operations.} To this end, KVMU ensures that streaming video LLMs can access critical cache data with low latency, maintain a reduced memory footprint, and utilize available bandwidth optimally through these two mechanisms.
 
\section{Evaluation}
\label{sec_eval}
\subsection{Experimental Setup}

\begin{table}[b]
\centering
\caption{Hardware Specifications of GPUs and V-Rex}
\includegraphics[width=0.49\textwidth]{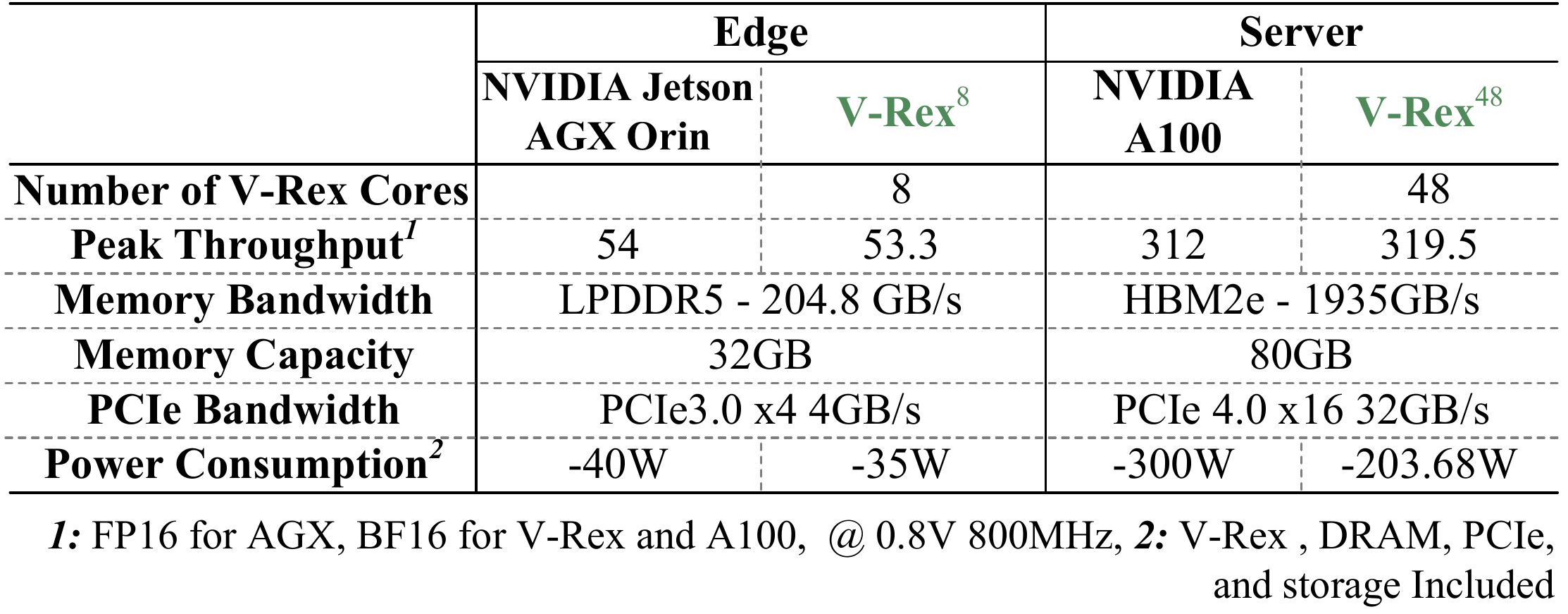}
\label{t1}
\end{table}

\begin{figure*}[t]
\centering
\includegraphics[width=\textwidth]{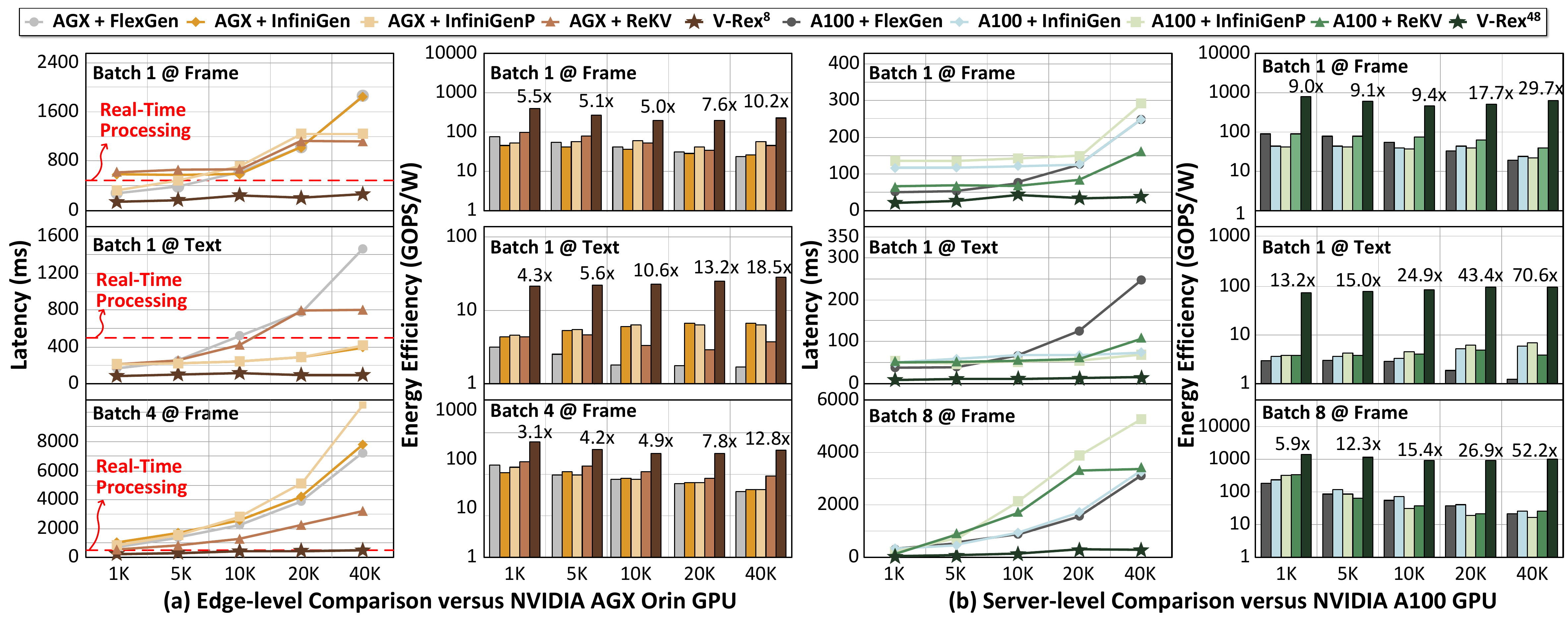}
\caption{Per-frame and TPOT latency and energy efficiency comparison versus (a) Edge GPU and (b) Server GPU. Baseline methods of FlexGen, InfiniGen, InfiniGenP, and ReKV are applied. We sweep the KV cache sequence length from 1K to 40K.}
\label{f12}
\end{figure*}

\textbf{Performance.} 
To evaluate the performance of \sysnames hardware system, we developed a custom cycle-level simulator. For DRAM performance, we integrated DRAMSim3~\cite{li2020dramsim3}, a widely used DRAM simulator, and for SSD performance, we incorporated MQSim~\cite{tavakkol2018mqsim}, an SSD simulator. To accurately model data movement between CPU memory and GPU memory, the actual data transfer bandwidth is modeled using an NVIDIA A100 GPU~\cite{nvidiaA100} and an AGX Orin GPU~\cite{nvidia_jetson_orin_2024}, both of which are incorporated into the simulator. We compared \sysname against two representative GPUs—an edge device (Jetson AGX Orin) and a server GPU (NVIDIA A100)—using identical system and performance parameters, summarized in Table~\ref{t1}. For the edge scenario, \sysname was instantiated with eight cores, utilizing the 4 GB/s PCIe with M.2 NVMe SSD for offloading the KV cache \textcolor{black}{and 204.8 GB/s LPDDR5 of 256-bit bus}. For the server scenario, \sysname utilized 48 cores, achieving a total of 319 TFLOPS, \textcolor{black}{with 1935 GB/s HBM2e of 5120-bit bus} and 32 GB/s PCIe with offloading the KV cache to DDR4-based CPU memory. For the streaming video LLM, all experiments employ Llama-3 8B as the backbone model and SigLIP-ViT-L-384~\cite{Zhai_2023_ICCV} as the vision encoder.

\textbf{Power/Area.} A single \sysname core is configured as $N_{DPE-h}$=64, $N_{DPE-w}$=64, $N_{VPE-h}$=1, $N_{VPE-w}$=64, $N_{HCU-h}$=1, $N_{HCU-w}$=16, $N_{WTU-h}$=1, and $N_{WTU-w}$=16. It was implemented in RTL and synthesized using Synopsys Design Compiler on a 14nm process. It operates reliably at 0.8 V and 800 MHz without timing violations \hypertarget{target:E_4_1}{}\textcolor{black}{under nominal conditions, as confirmed by pre-layout static timing analysis.} DRAM (HBM2e, DDR4) behavior was modeled using DRAMSim3, and LPDDR5 energy data were taken from vendor reports~\cite{kim2016future, lee2017understanding}. PCIe power was estimated at 3 W per lane under full load, and SSD power was based on Kioxia BG6 specifications~\cite{kioxia_bg6_2025}. GPU power measurements were obtained via NVIDIA-SMI and tegrastats~\cite{nvidia_smi_2008, nvidia_driveos_2019}. All these parameters were integrated into our custom simulator for accurate system-level evaluation. This setup ensures a realistic and fair comparison against baseline edge and server GPUs.

\subsection{Performance and Efficiency Evaluation}
\textbf{Latency.} To evaluate \sysname’s performance for streaming video LLMs, we compared its latency in frame processing and text generation against four top-k-based retrieval methods on both edge and server GPUs. FlexGen~\cite{sheng2023flexgen} serves as the baseline, offloading KV caches to CPU memory (A100) or storage (AGX Orin). InfiniGen~\cite{lee2024infinigen} retrieves tokens only during generation, InfiniGenP extends this to prefill, and ReKV~\cite{di2025streaming} performs frame-level selection. All baselines conduct KV prediction in the previous attention layer to prefetch KV caches, overlapping fetch latency with computation. We varied KV cache sizes (1K, 5K, 10K, 20K, 40K) using COIN~\cite{tang2019coin}, calibrating each method’s selection ratio to match baseline accuracy. Latency was measured as per-frame latency during frame processing and time per output token (TPOT) during text generation.

Latency comparison on the edge GPU is shown in Figure~\ref{f12} (a).
As token length increases, per-frame latency and TPOT rise across all prior methods due to heavier attention computation, greater selection overhead, and larger KV transfers, driven by fixed top-k requiring high token selection ratios. Consequently, none of the edge GPU setups—AGX+FlexGen, InfiniGen, InfiniGenP, or ReKV—achieve real-time performance at longer sequences, with the gap widening as token length grows. In the frame processing stage, the abundance of \textit{Query} tokens demands higher retrieval ratios than in text generation, since each query token requires retrieval. AGX+InfiniGen and AGX+InfiniGenP are even slower than the FlexGen baseline due to fine-grained, token-level selection introducing significant preprocessing overhead. AGX+ReKV’s coarse, frame-level selection offers modest latency gains but still requires selecting many tokens to maintain accuracy, limiting its effectiveness.

In contrast, V-Rex$^{8}$ enables real-time streaming inference ($\geq$2 FPS) even with long sequences and large batches. With a batch size of 1, per-frame latencies are 121 ms, 123 ms, 198 ms, 200 ms, and 254 ms for cache sizes of 1K, 5K, 10K, 20K, and 40K, respectively. It maintains 3.9–8.3 FPS across all sizes for real-time edge deployment, achieving 2.2–7.3$\times$ speedups over AGX+FlexGen. When the batch size increases to 4, speedups rise to 2.1–13.8$\times$. In text generation, TPOT latencies are lower, 89 to 97 ms, yielding 1.9–15.1$\times$ speedups. These gains stem from minimizing selected KV volume via \algoname and leveraging DRE’s high-speed compute and data movement. To evaluate scalability, we tested V-Rex$^{48}$ and an A100 GPU for server-level comparison (Figure~\ref{f12} (b)). \sysname achieves 20–48 ms per-frame latency, with 2.6–7.3$\times$ speedups at batch size 1. At batch size 8, speedups increase to 3.4–19.7$\times$, demonstrating strong parallel efficiency. TPOT latencies of 14–15 ms yield 2.8–16.8$\times$ speedups. These results show that \sysname significantly reduces latency in both frame processing and text generation for streaming video LLMs over edge and server GPUs.

\begin{figure}[t]
\centering
\hypertarget{target:C_1_2}{}
\includegraphics[width=0.49\textwidth]{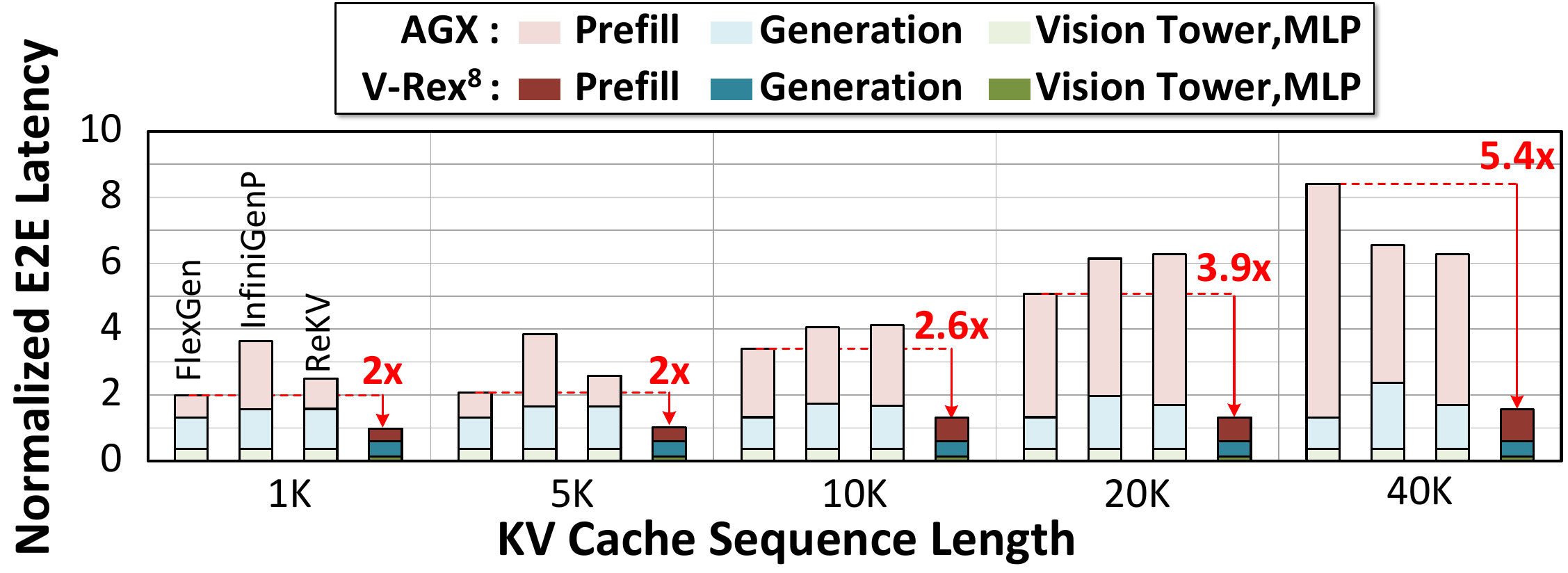}
\caption{\textcolor{black}{Comparison of End-to-End Latency Breakdown}}
\label{f_rc}
\end{figure}

\textcolor{black}{\textbf{E2E Latency Breakdown.} As shown in Figure~\ref{f_rc}, we evaluated the latency breakdown of AGX Orin and V-Rex$^{8}$ in an end-to-end streaming video LLM scenario, using an average case from the COIN benchmark. The results demonstrate that AGX+FlexGen fails to mitigate this explosive growth, as well as software-only optimizations (i.e., InfiniGenP and ReKV), which even perform slower than FlexGen from 1K to 20K due to KV prediction overhead. On the other hand, the primary performance gain of our work stems from reducing the overhead of the iterative prefill stage, increasing the performance gap as the KV cache sequence increases. This results in a reduction of up to 5.4$\times$ in end-to-end latency. By effectively managing the KV cache during prefill, our method maintains a consistent latency even as the cache grows.}

\textbf{Energy Efficiency.} Figure~\ref{f12} shows that \sysnames energy efficiency gains grow with token length, thanks to reduced data transfer. With batch size 1 during frame processing, \sysname achieves 5.5–10.2$\times$ greater energy efficiency over AGX+FlexGen for KV cache sizes from 1K to 40K; with batch size 4, the gain increases to 3.1–12.8$\times$. In text generation, the improvement is even more pronounced, reaching 4.3–18.5$\times$. This advantage is amplified on server GPUs, where power consumption is higher. Compared to A100+FlexGen, \sysname achieves 9.0–29.7$\times$ higher energy efficiency during frame processing with batch size 1. At batch size 8, it delivers 1.1–1.4 TOPS/W, achieving 5.9–52.2$\times$ gains. In text generation, energy efficiency reaches 13.2–70.6$\times$. These improvements stem from two key factors: \algoname’s ability to minimize retrieved data volume, and the KVMU module’s support for high-bandwidth, efficient data fetching. As a result, energy consumption for PCIe-based data transfers is significantly reduced. Overall, \sysname delivers substantially higher energy efficiency than state-of-the-art GPU-based retrieval methods, making it a compelling solution for streaming video LLM acceleration.

\begin{figure}[t]
\centering
\includegraphics[width=0.49\textwidth]{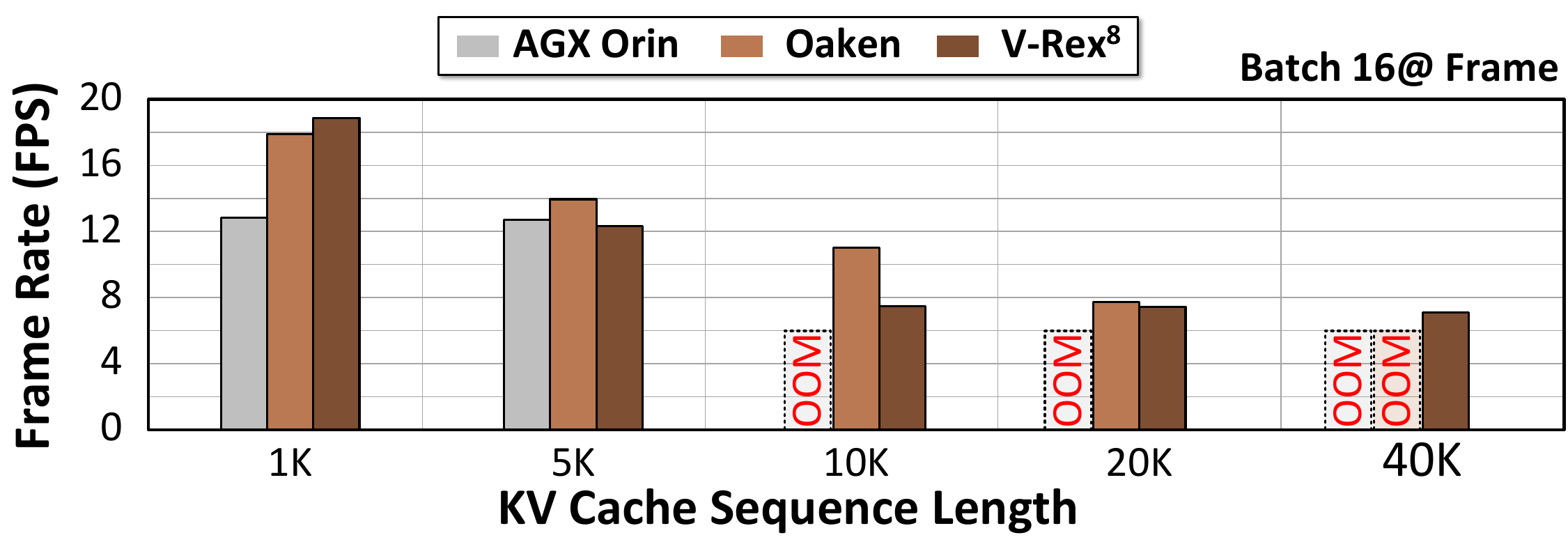}
\caption{Throughput Comparison versus SOTA LLM Accelerator}
\label{f13}
\end{figure}

\textbf{Comparison with SOTA Accelerator.} Figure~\ref{f13} compares the throughput of V-Rex$^{8}$ and Oaken~\cite{kim2025oaken}, a state-of-the-art LLM accelerator using 4-bit KV cache quantization, on the NVIDIA AGX Orin GPU. In this setup, AGX Orin runs FlexGen without KV offloading, while Oaken applies online quantization. At a short sequence length (1K), \sysname achieves up to 1.5$\times$ and 1.1$\times$ higher FPS than AGX Orin and Oaken, respectively, due to fully overlapped storage fetches and reduced attention computation. As sequence length increases, AGX Orin encounters out-of-memory (OOM) errors first, driven by the growing KV cache. Oaken, with its quantized cache, handles longer sequences and maintains higher throughput, but still fails beyond 20K tokens due to memory limits. In contrast, \sysnames efficient retrieval allows it to operate reliably beyond 20K tokens, sustaining 7 FPS even at large sequence lengths.

\begin{figure}[t]
\centering
\includegraphics[width=0.49\textwidth]{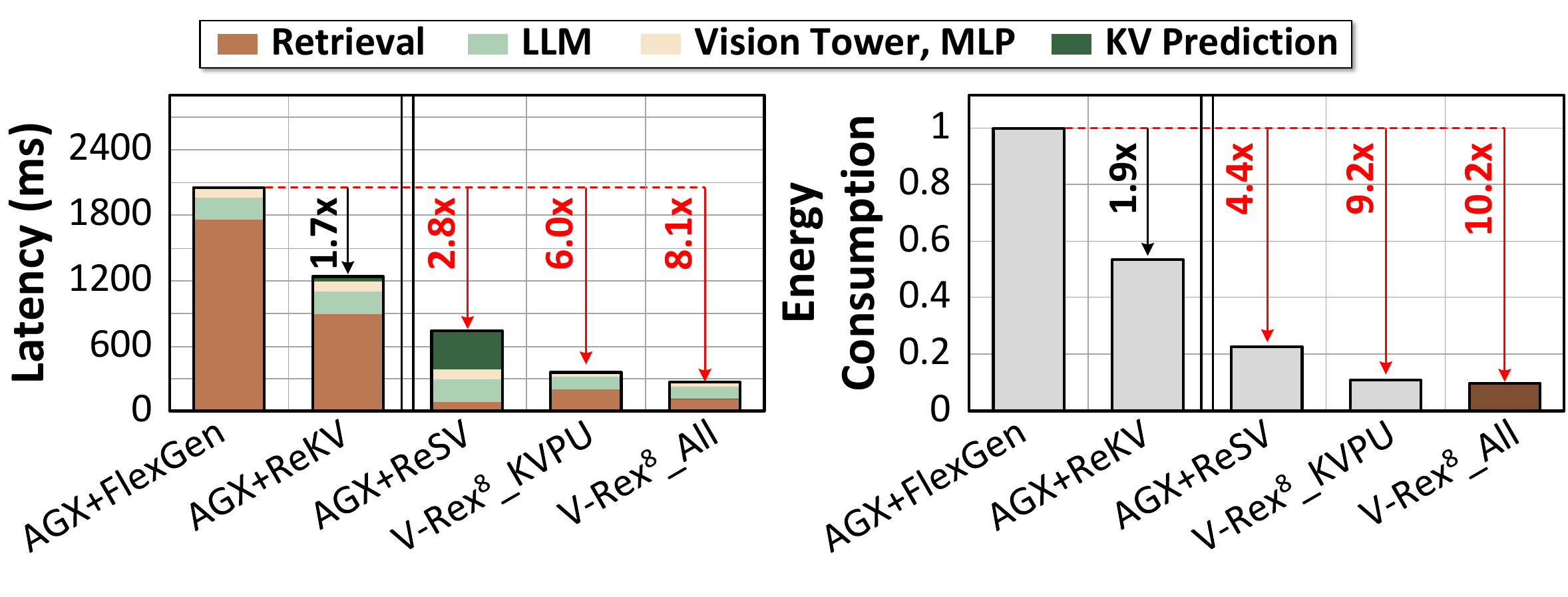}
\caption{Ablation Study and Latency Breakdown of \sysname}
\label{f14}
\end{figure}

\textbf{Ablation Study \& Latency Breakdown.} This evaluation illustrates how each \sysname optimization contributes to reducing latency and energy consumption during frame processing. It first presents cumulative gains as each optimization is applied, followed by a latency breakdown showing which execution components are affected by each scheme. We implemented AGX+ReSV by applying \algoname on the AGX Orin GPU and evaluated V-Rex$^{8}$ by incrementally enabling optimizations under a 40K cache with batch size 1. V-Rex$^{8}$\_KVPU includes the KVPU, while V-Rex$^{8}$\_All incorporates all optimizations, including KVMU. The results clearly demonstrate the GPU’s inefficiency and highlight the need for software-hardware co-design.

As shown in Figure~\ref{f14}, AGX+ReSV reduces overall latency by 2.8$\times$ by hiding most retrieval overhead under computation. However, due to complex KV prediction, such as conditional and data-dependent computation for clustering and thresholding, it still accounts for 48\% of total latency, limiting the algorithm’s full potential on GPU. With hardware-level optimization, V-Rex$^{8}$\_KVPU reduces KV prediction latency overhead down to 0.5\% (from 23\% of total computation), achieving a 6.0$\times$ speedup and 9.2$\times$ energy reduction by overlapping prediction operation with LLM computation using HCU’s fast bit-wise operations and WTU’s early-exit sorting. V-Rex$^{8}$\_All further improves performance by increasing PCIe bandwidth utilization, reaching an 8.1$\times$ speedup and 10.2$\times$ energy savings. Although KVMU introduces some memory overhead due to the HC table, it occupies only 1.67\% of the full KV cache with an average of 32 tokens per cluster. Each \sysname optimization contributes incrementally to performance and energy efficiency. Notably, \algoname alone is insufficient; the combined effect of \algoname and DRE is essential to fully realize efficient KV cache retrieval for streaming video LLMs.

\begin{figure}[t]
\centering
\includegraphics[width=0.5\textwidth]{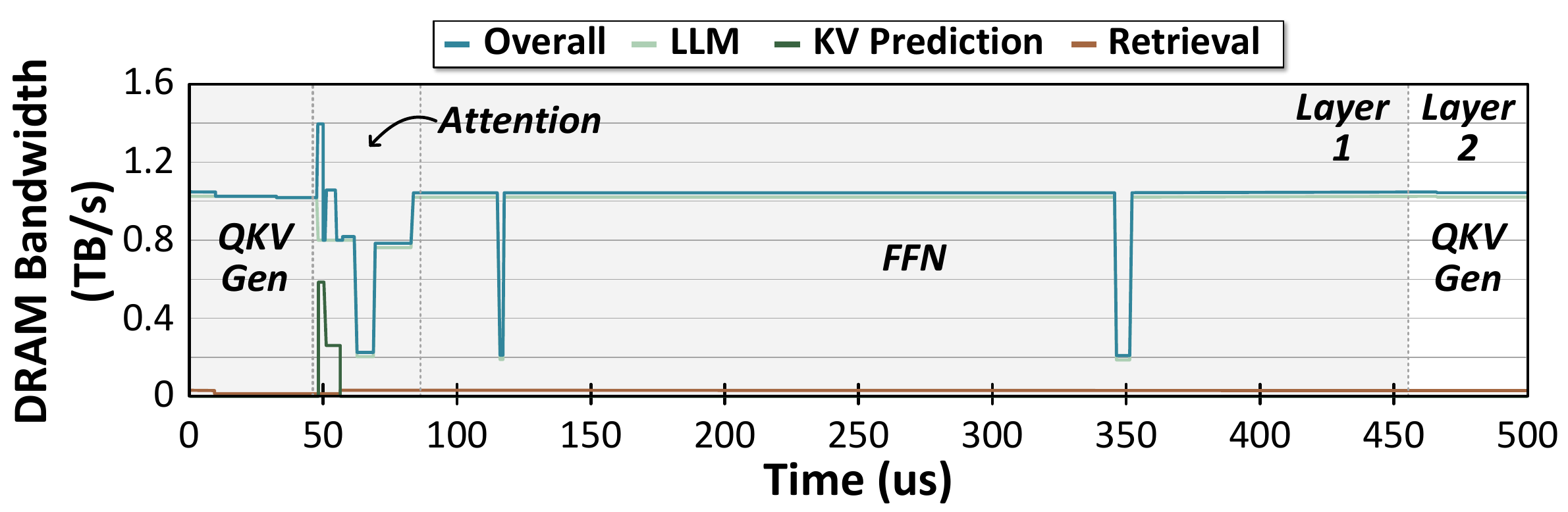}
\caption{Anaylsis on Memory Bandwidth Usage of V-Rex$^{48}$}
\label{f_re}
\end{figure}

\begin{figure}[t]
\centering
\includegraphics[width=0.49\textwidth]{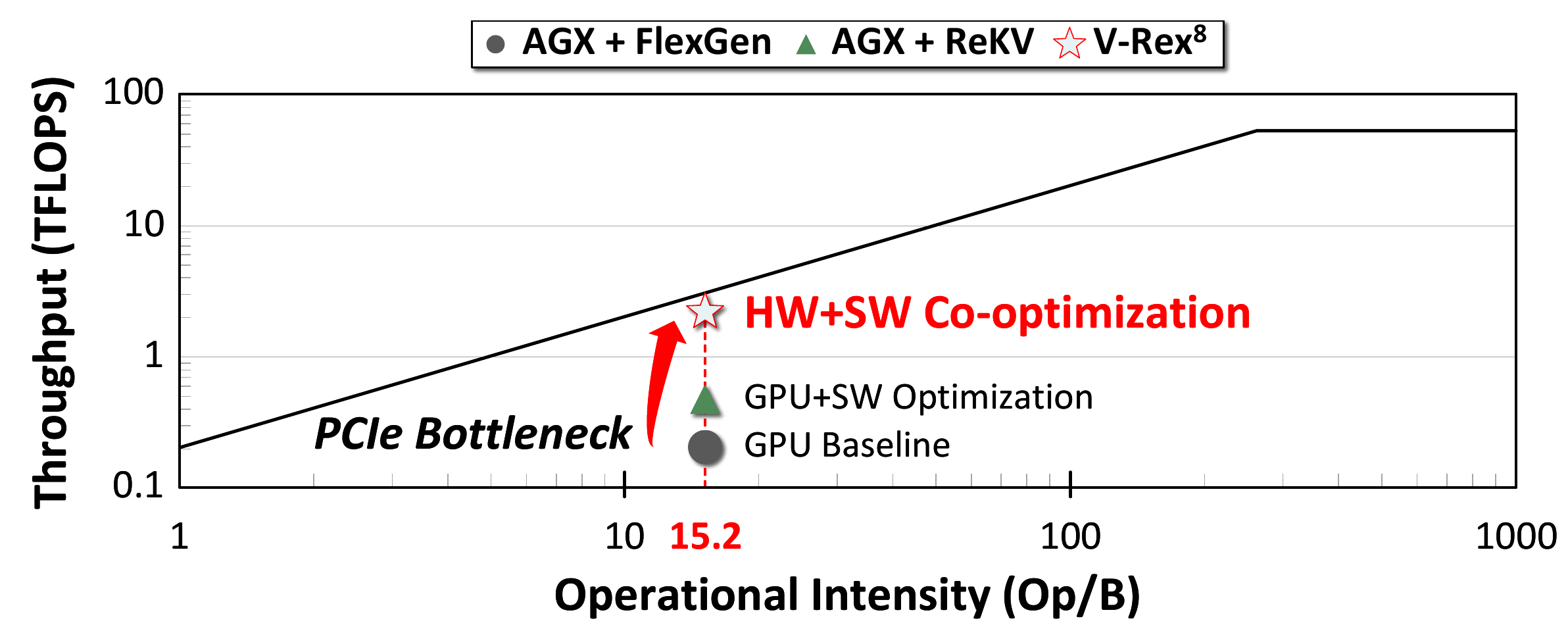}
\caption{\textcolor{black}{Roofline Model Analysis on AGX Orin and V-Rex$^{8}$}}
\label{f_re5}
\end{figure}

\subsection{Bandwidth Analysis for Concurrent Computation}
To show that KV prediction and retrieval can be fully overlapped with LLM computations, we analyzed the bandwidth usage of each computation over a layer of frame processing stage, as shown in Figure~\ref{f_re}. It demonstrates that memory is effectively shared among concurrent operations. The KV prediction is executed concurrently with the attention operation. Although it briefly spikes bandwidth usage to 600GB/s, its short duration allows it to be hidden entirely. The KV retrieval, which transfers data from CPU memory to DRAM, runs for most of the execution time but only consumes about 1\% of the bandwidth. Because KV cache fetch is bottlenecked by PCIe bandwidth, which is about 1\% of DRAM bandwidth, it allows KV retrieval to be concurrently executed with attention and FFN computations with minimal overhead.

\subsection{\textcolor{black}{Roofline Model Analysis}}
\textcolor{black}{Figure~\ref{f_re5} illustrates a roofline model analysis of the frame processing stage for three edge-level systems: AGX+FlexGen, AGX+ReKV, and our proposed V-Rex$^{8}$. This analysis uses a workload with a KV cache length of 40K and a batch size of 4, yielding an average operational intensity of 15.2 Op/B. The result reveals a significant disparity in the achieved throughput across the systems. AGX+FlexGen exhibits the lowest performance, reaching only 6.6\% of its theoretical maximum. This severe underutilization is attributed to the slow PCIe communication, which creates a bottleneck during KV cache fetching. Therefore, optimizing the LLM inference computation alone is ineffective without resolving the fundamental I/O bottleneck. AGX+ReKV employs a retrieval mechanism to achieve a higher throughput, reaching approximately 15\% of the peak. However, being a purely software-based optimization, it remains inefficient. Finally, our proposed V-Rex demonstrates a remarkable throughput at 71.5\% of its theoretical maximum, marking a 10.8$\times$ improvement over AGX+FlexGen. It confirms that V-Rex effectively resolves the inefficiencies inherent in conventional GPU-based systems.}

\subsection{Comparative Accuracy Analysis}
\textbf{Workload.} To demonstrate the flexibility and accuracy of \algoname, we evaluated and compared the performance of existing retrieval methods (i.e., InfiniGen, InfiniGenP, and ReKV) using five benchmarks from the COIN dataset. VideoLLM-Online~\cite{chen2024videollm} was used as the baseline without any retrieval optimization applied. For this experiment, existing methods were configured to select up to 50\% of tokens with their fixed top-k mechanism, while \algoname used a threshold in its WiCSum operation that was empirically tuned to ensure the accuracy, configuring $N_{hp}$=32,
$Th_{wics}$ to 0.3 and $Th_{hp}$=7. 

\begin{table}[t]
\centering
\caption{Model Accuracy Evaluation and Retrieval Ratio}
\includegraphics[width=0.49\textwidth]{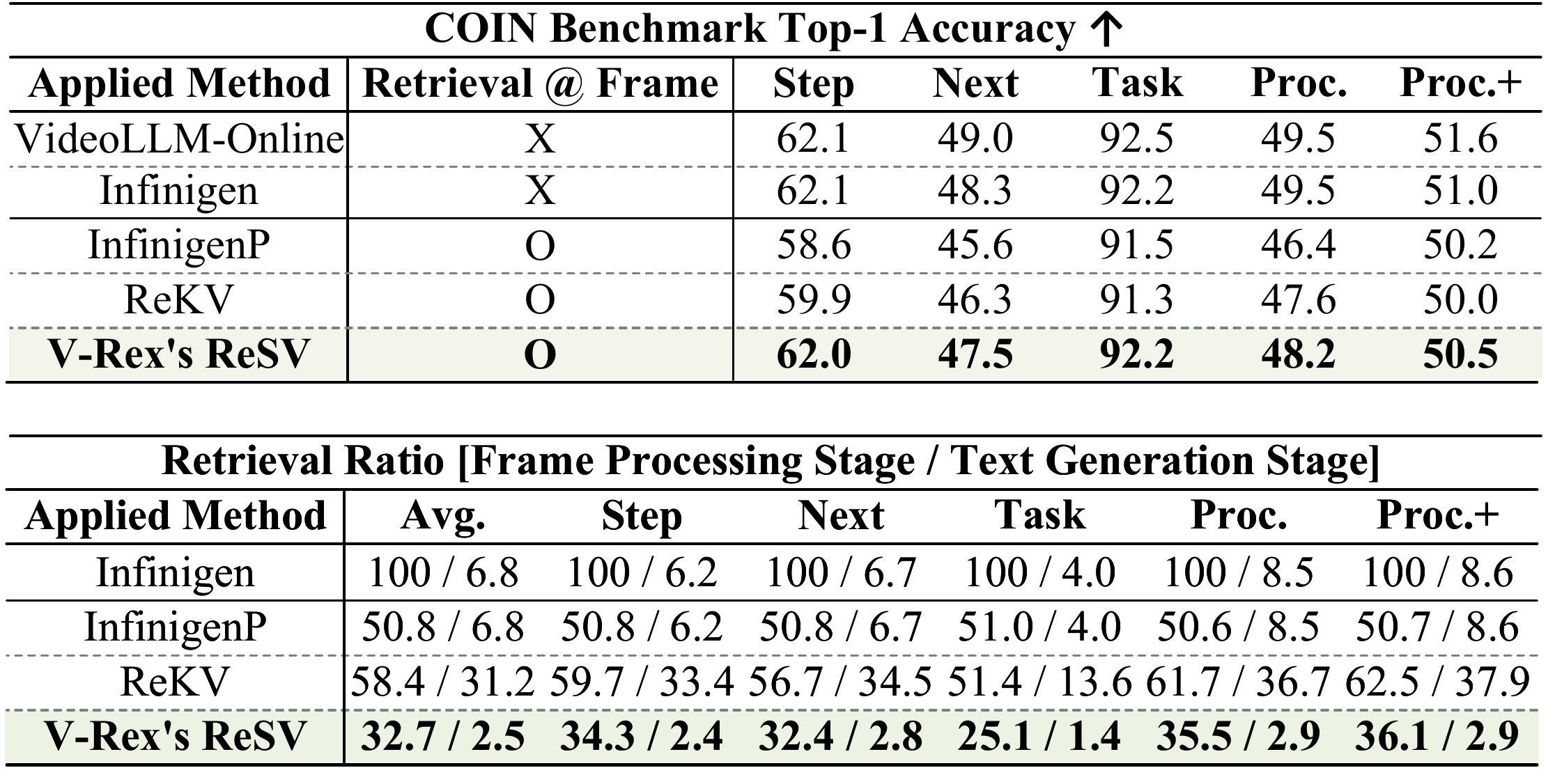}
\label{t2}\end{table}

\begin{figure}[t]
\centering
\includegraphics[width=0.49\textwidth]{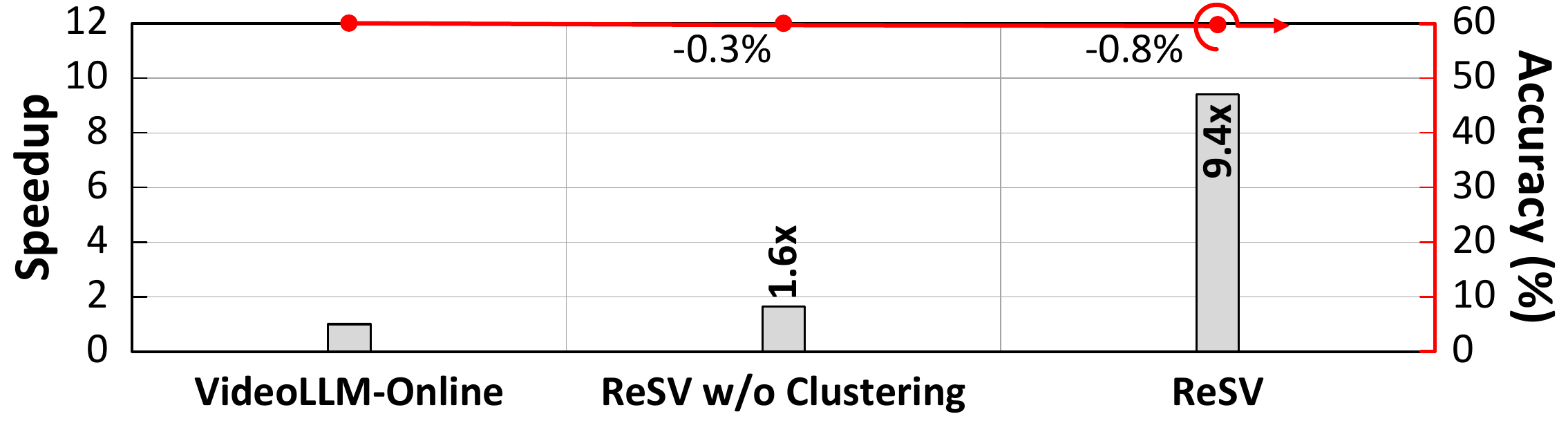}
\caption{Ablation Study of \algoname}
\label{fre2}
\end{figure}

\begin{figure}[t]
\centering
\includegraphics[width=0.49\textwidth]{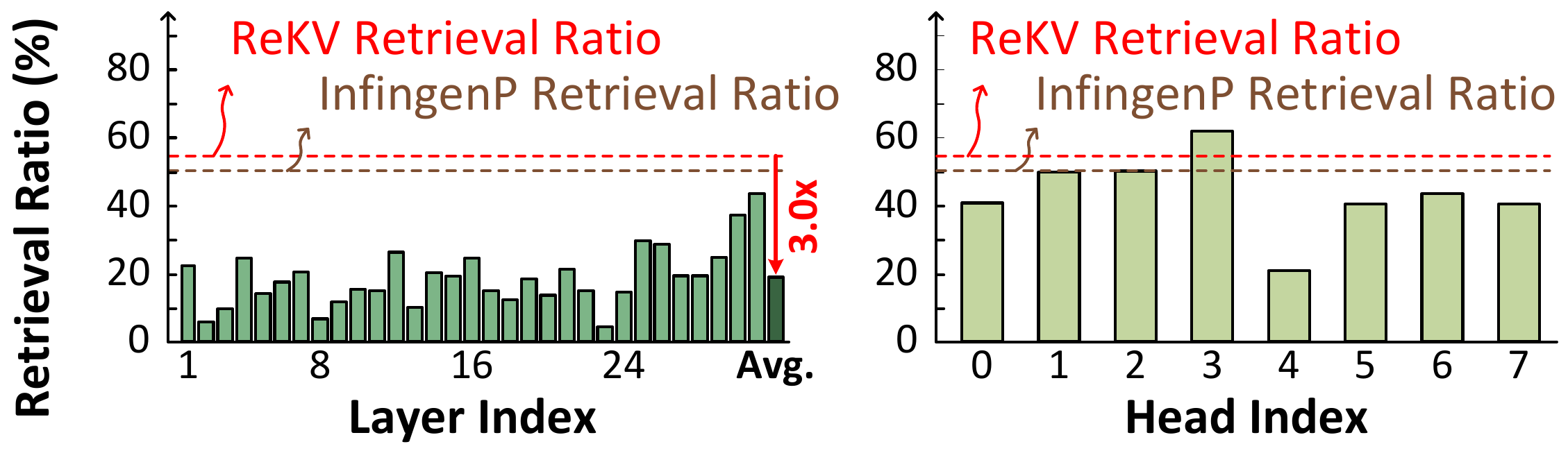}
\caption{Comparison of Retrieval Ratio per Layer and per Head}
\label{f15}
\end{figure}

\textbf{Accuracy.} Table~\ref{t2} summarizes the results. \sysnames \algoname outperforms other retrieval methods, demonstrating the lowest retrieval ratio while achieving the highest overall accuracy. Compared to the baseline vanilla model (VideoLLM-Online), \algoname exhibits only a marginal average accuracy drop of 0.8\%. Additionally, \algoname significantly reduces the retrieval ratio, as it can adopt diverse score distributions from various tasks. During the frame processing stage, the average retrieval ratio ranges from 25.1\% to 36.1\%, and during the text generation stage, it varies between 1.4\% and 2.9\%. This variability highlights that the thresholding mechanism in \algoname effectively adapts token selection according to each task’s characteristics.

In contrast, InfiniGen maintains accuracy comparable to the vanilla model, but it does not perform retrieval during the frame processing stage, making it impractical for real-time inference. InfiniGenP retrieves approximately 50\% of tokens during the frame processing stage, which leads to a substantial accuracy degradation of up to 3.4\%. ReKV, which employs frame-wise selection, requires a large volume of retrieved KV cache for both frame processing and text generation stages to maintain the accuracy as InfiniGenP, thus degrading the efficiency. In summary, the hash-bit key clustering and WiCSum thresholding techniques of \algoname enable dynamic adaptation to data distribution, effectively selecting the minimal number of tokens while preserving accuracy. This makes \algoname particularly suitable for real-time and resource-constrained streaming video LLM inference.

\textbf{\algoname Efficiency.} We performed an ablation study by incrementally applying ReSV's optimizations. Figure~\ref{fre2} shows the average accuracy on COIN benchmarks and the frame processing latency at 40K cache size. First, ReSV without applying clustering improves latency by 1.6$\times$ over the baseline, causing a minor accuracy drop of 0.3\%, originating from the light attention computation. Second, ReSV, which further incorporates hash-bit clustering, achieves a 9.4$\times$ speedup, accompanied by a 0.8\% accuracy reduction. This significant speedup comes from reducing the fetching and computing of the entire key for WiCSum thresholding computation by clustering the key cache.

Figure~\ref{f15} presents the ratio of retrieved data on a per-layer and per-head basis of a sample video from COIN. Unlike InfiniGenP and ReKV, which retrieve a fixed number of KV cache tokens uniformly across all layers and heads, \algoname exhibits a diverse distribution in the token retrieval ratio. Certain layers that require fewer tokens exhibit selection rates of 4.2\%, while more critical layers with higher token importance demonstrate around 44.0\%. This variability can also be observed among the attention heads. It enables \algoname to maintain higher accuracy while retrieving 3.0$\times$ fewer tokens on average compared to ReKV, resulting in superior efficiency compared to fixed top-k mechanisms.

\hypertarget{target:B_5_1}{}\subsection{\sysnames Hardware Overhead} 
\textbf{Power and Area.} Table~\ref{t3} summarizes the power consumption and area breakdown for \sysname equipped with a single core. A single \sysname core occupies 1.89 mm$^{2}$ and consumes 2.61 W, \textcolor{black}{equipped with on-chip memory of 384 KB for LXE and 20.125 KB for DRE.} When scaled to larger configurations, the area of V-Rex$^{8}$ is 15.12 mm$^{2}$, which is substantially smaller than the AGX Orin GPU (200 mm$^{2}$). Notably, V-Rex$^{48}$ occupies 90.57 mm$^{2}$, considerably less than the A100 GPU (826 mm$^{2}$). Including overall system power, V-Rex$^{8}$ consumes 35 W, achieving 11.4\% lower power consumption than the AGX Orin GPU (40 W), while V-Rex$^{48}$ consumes 203.68~W, demonstrating 32.1\% lower power consumption than the A100 GPU (300~W), as detailed in Table~\ref{t1}. The additional hardware overhead of DRE is minimal, accounting for only 2.4\% of the chip's total power and 2.0\% of the total area, which can be attributed to the effective KV cache retrieval algorithm. Its compact design enables efficient integration with any existing GPUs, NPUs, and LLM accelerators.

\begin{table}[t]
\centering
\caption{Breakdown of Area and Power}
\includegraphics[width=0.49\textwidth]{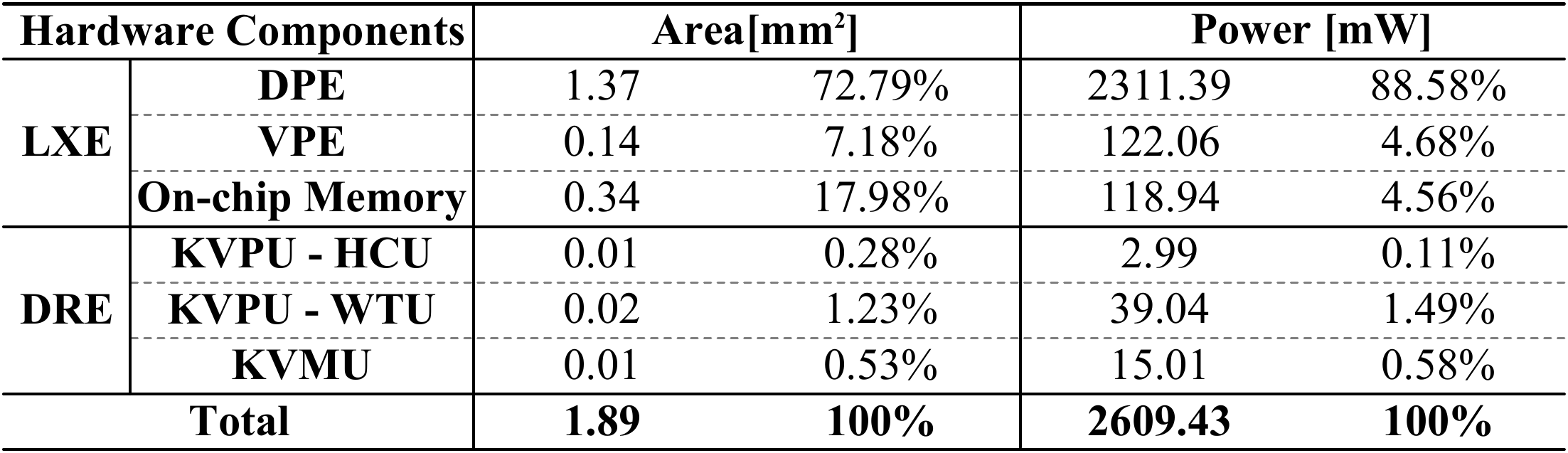}
\label{t3}
\end{table}   
\section{Related Work \& Discussion}
\label{sec_discussion}

\textbf{Streaming Video LLM Optimization.} VideoLLM-Online~\cite{chen2024videollm} introduces a streaming video interaction system that adjusts frame sampling and resolution. LiveVLM~\cite{ning2025livevlm} focuses on reusable short- and long-term memory tokens for efficient frame understanding. QuickVideo~\cite{schneider2025quickvideo} leverages parallel CPU decoding and GPU inference overlap to achieve end-to-end speedups. VidMoD~\cite{wu2024videollm} reduces visual processing workload using mixture-of-depth computation, dynamically skipping unnecessary layers. While these systems improve frame processing efficiency, they still suffer from repeated KV cache transfers and memory overhead during multi-query streaming workloads. In contrast, \sysname integrates clustering and dynamic thresholding directly into the prefill stage to manage KV transfers more effectively.

\textbf{KV Cache Management and Compression.} To address the growing memory cost of KV caches, various methods reduce KV size or selectively fetch relevant tokens. LeanKV~\cite{zhang2025unifyingkvcachecompression} employs multi-precision KV cache quantization, where high-precision keys and low-precision values are used. MiniCache~\cite{10.5555/3737916.3742359} and PyramidInfer~\cite{yang2024pyramidinfer} compress based on layer-wise importance but leave inter-frame similarity unexplored. CacheGen~\cite{CacheGen} exploits a traditional video codec mechanism but requires expensive pre-processing overhead for compression, which is not suitable for streaming applications. All these methods use fixed compression and pruning policies that cannot adapt to the evolving attention patterns in video streams. \algoname clusters similar tokens across frames, and its dynamic token selection per layer and head yields more precise retrieval without sacrificing context integrity.

\textbf{Transformer Hardware Accelerators.} Previous works primarily focused on computational and memory bandwidth optimizations in accelerating LLMs. DFX~\cite{hong2022dfx} and SpecEE~\cite{xu2025specee} target to accelerate the memory-bound generation stage. BitMoD~\cite{chen2025bitmod} and LUT Tensor Core~\cite{mo2024lut} reduce computational overhead with a low-bit inference method. While AiF~\cite{lee2025aif} demonstrates in-flash processing for on-device LLMs, it does not incorporate dynamic retrieval mechanisms. Oaken~\cite{kim2025oaken} introduces KV cache quantization, which effectively increases the maximum cache capacity but does not address the fundamental structural issue of OOM errors caused by unbounded cache growth. Our method can be applied on top of such prior techniques to further optimize streaming video LLM inference and tackle challenges beyond simple cache size expansion. For vision-oriented models, accelerators such as Adaptiv~\cite{yoo2024adaptiv} and EXION~\cite{heo2025exion} primarily focus on improving compute density. However, these designs do not directly address the growth of KV cache overhead unique to streaming video LLM.

\section{Conclusion}
We presented \sysname, the first end-to-end accelerator and KV cache management solution tailored for streaming video LLMs with dynamic KV cache retrieval. Our contributions span the algorithm and hardware levels. Through \algoname, a training-free, dynamic KV retrieval method, \sysname reduces KV cache volume with negligible accuracy drop. To support ReSV efficiently, we designed a compact hardware unit, DRE, supporting low-latency and energy-efficient computation. \sysname enables real-time streaming inference, achieving 3.9-8.3~FPS on an edge deployment with 1.9-19.7$\times$ speedup and 3.1-18.5$\times$ energy efficiency gains, and extends to 2.6-19.7$\times$ speedup gains and 5.9-70.6$\times$ energy efficiency gains over a server GPU. As KV cache retrieval becomes increasingly crucial in long-context and streaming LLMs, we believe \sysname presents a promising direction for future research in real-time, energy-efficient LLM acceleration, particularly for resource-constrained edge environments and scalable deployment.

\section*{Acknowledgements}

This work was supported by Institute for Information \& communications Technology Promotion (IITP) grant funded by the Korea government (MSIT) (No. RS-2025-02264029, Integration and Validation of an AI Semiconductor-Based Data Center Training and Inference System and No. IITP-2025-RS-2023-00256472, Graduate School of Artificial Intelligence Semiconductor).

\bibliographystyle{IEEEtranS}
\bibliography{refs}

\end{document}